\documentclass[draft]{agujournal2019}
\usepackage{url} %this package should fix any errors with URLs in refs.
\usepackage{lineno}
\usepackage[inline]{trackchanges} %for better track changes. finalnew option will compile document with changes incorporated.
\usepackage{soul}
\draftfalse

\journalname{Space Weather}

\begin{document}

%% ------------------------------------------------------------------------ %%
%  Title
%
% (A title should be specific, informative, and brief. Use
% abbreviations only if they are defined in the abstract. Titles that
% start with general keywords then specific terms are optimized in
% searches)
%
%% ------------------------------------------------------------------------ %%

% Example: \title{This is a test title}

\title{Collection, Collation, and Comparison of 3D Coronal CME Reconstructions}

%% ------------------------------------------------------------------------ %%
%
%  AUTHORS AND AFFILIATIONS
%
%% ------------------------------------------------------------------------ %%

% Authors are individuals who have significantly contributed to the
% research and preparation of the article. Group authors are allowed, if
% each author in the group is separately identified in an appendix.)

% List authors by first name or initial followed by last name and
% separated by commas. Use \affil{} to number affiliations, and
% \thanks{} for author notes.
% Additional author notes should be indicated with \thanks{} (for
% example, for current addresses).

% Example: \authors{A. B. Author\affil{1}\thanks{Current address, Antartica}, B. C. Author\affil{2,3}, and D. E.
% Author\affil{3,4}\thanks{Also funded by Monsanto.}}

\authors{C.~Kay\affil{1,2}~and~E.~Palmerio\affil{3}}

\affiliation{1}{Heliophysics Science Division, NASA Goddard Space Flight Center, Greenbelt, MD, USA}
\affiliation{2}{Dept.\ of Physics, The Catholic University of America, Washington, DC, USA}
\affiliation{3}{Predictive Science Inc., San Diego, CA, USA}

%% Corresponding Author:
% Corresponding author mailing address and e-mail address:

% (include name and email addresses of the corresponding author.  More
% than one corresponding author is allowed in this LaTeX file and for
% publication; but only one corresponding author is allowed in our
% editorial system.)

% Example: \correspondingauthor{First and Last Name}{email@address.edu}

\correspondingauthor{Christina Kay}{christina.d.kay@nasa.gov}

%% Keypoints, final entry on title page.

%  List up to three key points (at least one is required)
%  Key Points summarize the main points and conclusions of the article
%  Each must be 140 characters or fewer with no special characters or punctuation and must be complete sentences

% Example:
% \begin{keypoints}
% \item	List up to three key points (at least one is required)
% \item	Key Points summarize the main points and conclusions of the article
% \item	Each must be 140 characters or fewer with no special characters or punctuation and must be complete sentences
% \end{keypoints}

\begin{keypoints}
\item LLAMACoRe combines 24 different catalogs of 3D CME reconstructions and identifies which entries correspond to the same event
\item We determine the most likely uncertainties in the reconstructed latitude, longitude, tilt, angular width, shape, velocity, and mass 
\item Most events are well-described by the most likely uncertainty but we do find a few cases with extremely-differing reconstructions
\end{keypoints}

%% ------------------------------------------------------------------------ %%
%
%  ABSTRACT and PLAIN LANGUAGE SUMMARY
%
% A good Abstract will begin with a short description of the problem
% being addressed, briefly describe the new data or analyses, then
% briefly states the main conclusion(s) and how they are supported and
% uncertainties.

% The Plain Language Summary should be written for a broad audience,
% including journalists and the science-interested public, that will not have 
% a background in your field.
%
% A Plain Language Summary is required in GRL, JGR: Planets, JGR: Biogeosciences,
% JGR: Oceans, G-Cubed, Reviews of Geophysics, and JAMES.
% see http://sharingscience.agu.org/creating-plain-language-summary/)
%
%% ------------------------------------------------------------------------ %%

%% \begin{abstract} starts the second page

\begin{abstract}
Predicting the impacts of coronal mass ejections (CMEs) is a major focus of current space weather forecasting efforts. Typically, CME properties are reconstructed from stereoscopic coronal images and then used to forward model a CME's interplanetary evolution. Knowing the uncertainty in the coronal reconstructions is then a critical factor in determining the uncertainty of any predictions. A growing number of catalogs of coronal CME reconstructions exist, but no extensive comparison between these catalogs has yet been performed. Here we develop a Living List of Attributes Measured in Any Coronal Reconstruction (LLAMACoRe), an online collection of individual catalogs, which we intend to continually update. In this first version, we use results from 24 different catalogs with 3D reconstructions using STEREO observations between 2007--2014. We have collated the individual catalogs, determining which reconstructions correspond to the same events. LLAMACoRe contains 2954 reconstructions for 1862 CMEs. Of these, 511 CMEs contain multiple reconstructions from different catalogs. Using the best-constrained values for each CME, we find that the combined catalog reproduces the generally known solar cycle trends. We determine the typical difference we would expect between two independent reconstructions of the same event and find values of 4.0$^{\circ}$ in the latitude, 8.0$^{\circ}$ in the longitude, 24.0$^{\circ}$ in the tilt, 9.3$^{\circ}$ in the angular width, 0.1 in the shape parameter $\kappa$, 115~km/s in the velocity, and 2.5$\times$10$^{15}$~g in the mass. These remain the most probable values over the solar cycle, though we find more extreme outliers in the deviation toward solar maximum.
\end{abstract}

\begin{plainlanguagesummary}
Coronal mass ejections (CMEs) are large explosions from the solar atmosphere that propagate out through space. Knowing where they are going is important for predicting space weather. We have satellites staring at the Sun from different directions, which we can use to reconstruct the 3D location and orientation of a CME. A few groups have routinely reconstructed nearly every CME over long time spans, other groups reconstruct tens of events for smaller, focused studies. Here we collect as many reconstructions as we can find for CMEs that occurred between 2007--2014. We match the reconstructions between catalogs, producing a new catalog, LLAMACoRe, that contains 1862 CMEs with at least one reconstruction. 511 CMEs have more than one reconstruction. We compare the behavior of this set to previously known trends and find good general agreement. For the cases with multiple reconstructions, we can quantify how much uncertainty we expect in the reconstructed values, which should be a useful metric for forecasters. We find that the most probable uncertainty in each parameter tends to remain relatively constant over time. When the Sun becomes more active and releases more energetic CMEs there are more outlier cases, but most cases remain at lower values.
\end{plainlanguagesummary}

%% ------------------------------------------------------------------------ %%
%
%  TEXT
%
%% ------------------------------------------------------------------------ %%

\section{Introduction}
Coronal mass ejections (CMEs) are known to be major drivers of space weather effects at Earth as well as elsewhere in the solar system and, as such, they are routinely monitored via remote-sensing imagery and included in forecasting models by space weather agencies \cite<e.g.,>{Piz11, Sha17}. When a CME is ejected from the Sun, its eruptive signatures are usually observed on the solar disk and/or off the limb in extreme ultraviolet or X-ray data \cite<e.g.,>{Hud01}, and its evolution through the corona is often followed in white-light coronagraph images \cite<e.g.,>{Vou13}. While some models employed in heliophysics research are able to simulate the eruption and evolution of CMEs from the lower solar atmosphere outwards \cite<using photospheric data as a boundary condition; e.g.,>{Jin17, Kay22, Lyn21, Tor18}, many research models and all operational ones avail themselves of coronagraph data to derive input parameters to inject CMEs into the computational domain at some altitude in the outer corona \cite<often set at 21.5\,$R_{\odot}$ or 0.1~au; e.g.,>{Dum21, Mah22, Mos18, Ods23}. Even more so, in the former class of models white-light coronal imagery is frequently employed to validate simulation results \cite<e.g.,>{Kay17AR, Lug07, Lyn16, Man08}.

One of the major issues of observing CMEs in the corona remotely is that the line-of-sight integrated electron density measured by a white-light camera manifests itself as a 2D image. This means that most of the information on the 3D structure and trajectory of a CME is not retrieved, and the corresponding geometric and kinematic properties can only be analyzed based on their projection onto a plane. A major turning point in remote-sensing observations of CMEs was reached in 2006 with the launch of the twin Solar Terrestrial Relations Observatory \cite<STEREO;>{Kai08} spacecraft away from the Sun--Earth line, which enabled stereoscopic views of the solar corona alongside the near-Earth perspective afforded by the Solar and Heliospheric Observatory \cite<SOHO;>{Dom95} mission, launched in 1995 and located at the Sun--Earth Lagrange L1 point. Taking advantage of observations of the same CME from three viewpoints, the heliophysics research community has developed a wealth of models and techniques to recover the 3D structure of CMEs in the corona using multi-point, remote-sensing measurements. These methods include triangulation \cite<e.g.,>{Inh06, Lie11, Mie08, Mor10} and fitting \cite<e.g.,>{Isa16, Mil13, The09, Zha22} techniques that, while similar in practical implementation, differ in their basic assumptions about the CME morphology---i.e., fitting techniques require a pre-determined geometry whereas triangulation techniques focus on identifiable features in the data without requiring such models. Examples of different available reconstruction methods were reviewed and applied to a few events observed in coronagraph data from two or three viewpoints by \citeA{Mie10} and \citeA{Fen13}.

Due to their practicality of use, coronagraph fitting tools have been extensively included in CME analyses, e.g., to investigate the coronal evolution of case study events \cite<e.g.,>{Dum21wde, Vou11}, to derive input parameters for heliospheric modeling \cite<e.g.,>{May15, Pal19}, and even to interpret simulation data \cite<e.g.,>{Ben23, Ver23}. The basic principle behind white-light fitting techniques is that a parameterized CME 3D ``shape'' is projected onto nearly-simultaneous images (usually from two or three coronagraphs) and manually adjusted until it visually matches observations to the best of the user's abilities. Among the simplest CME geometries is the so-called Ice Cream Cone \cite<ICC;>{xue05b} model, which consists of a conical base culminating in a semi-spherical front and is fully defined by four properties: latitude, longitude, and height of the apex, as well as the width of the cone. A conical geometry is implemented, e.g., by the Stereo CME Analysis Tool \cite<StereoCAT;>{DONKI}, which uses imagery from two viewpoints and a cone's projection onto them to retrieve 3D CME parameters. A similar morphology is implemented in the ``teardrop'' 3D lemniscate employed by the NOAA Space Weather Prediction Center CME Analysis Tool \cite<SWPC-CAT;>{Mil13}. More complex geometries include the croissant-shaped Graduated Cylindrical Shell \cite<GCS;>{The11} model, which to the ICC parameters of latitude, longitude, and height of the CME apex adds the tilt of the axis, the distance between the two legs, and the aspect ratio of the parameterized shell. Further modifications to the GCS morphology include the Flux Rope in 3D \cite<FRi3D;>{Isa16} model, which adds three additional distortion parameters (skew, pancaking, and flattening) to reproduce a larger variety of CME characteristics in the corona. By fitting the same CME at two or more times using any of these models, it is possible to derive the speed of the apex (or nose). Since the pre-determined CME geometries for coronagraph fitting consist of ``hollow'' 3D shells, it is not possible to determine CME mass using such models alone, but they can be combined with other analysis methods \cite<e.g.,>{Col09, Plu19, Sav13} to ultimately yield a mass estimate.

Despite the ease of use and widespread availability of coronagraph fitting techniques in the research and operational communities, there are still several issues to be considered. First of all, many CMEs observed in the corona deviate significantly from the generally well-behaved geometries assumed by the models, making reconstructions particularly difficult \cite<e.g.,>{Col15, Nie22}. Additionally, alongside the intrinsic uncertainties attached to each model, it is important to mind the inherent subjectivity of the user performing the fitting \cite<e.g.,>{Sin22, Ver23}. In fact, it has been shown that, even when the exact same background solar wind conditions are used, coronagraphic reconstructions of CMEs performed by different users, or even by the same individual employing different CME geometries, can lead to significantly different CME arrival time predictions \cite<e.g.,>{Pal22enlil}. To mitigate the innate uncertainties and errors associated with a single set of coronal reconstructions performed by a single user, the research community has often modeled CME propagation in the heliosphere either by running ensemble simulations \cite<e.g.,>{Dum18, Kay18, Lee13, May15AT} or by considering ``clouds'' of synthetic observers around the target of interest \cite<e.g.,>{Asv21, Mah23, Pal23, Sco19}.

Given the premises outlined above, it follows that having a ``super-database'' of coronal reconstructions available to the community would be beneficial not only to determine the characteristic uncertainties more precisely via large statistics, but also for forecast-oriented researchers to test and validate their models. To this end, we have perused the existing literature and collected published catalogs of CME coronal reconstructions to the best of our abilities, making sure to group together fits performed on the same event. We have then analyzed the full ``meta-catalog'' in search of possible trends, including explorations of characteristic errors with respect to various CME properties and how such uncertainties vary with differing numbers of available reconstructions. The manuscript is organized as follows. In Section~\ref{sec:data}, we present the individual catalogs and specify how reconstructions have been connected to single CME events. In Section~\ref{gen}, we analyze and discuss the general CME properties (including distributions, variations, and correlations) emerging from the collected reconstructions. In Section~\ref{multiSec}, we focus on the events that appear in two or more catalogs and determine the characteristic uncertainties and their variation with number of reconstructions. We discuss our results in Section~\ref{disc} and present our conclusions in Section~\ref{sec:conc}.

\section{Data Collection} \label{sec:data}
We have created the largest-to-date collection of 3D coronal CME reconstructions by combining the results of existing catalogs. This Living List of Attributes Measured in Any Coronal Reconstruction, or LLAMACoRe, is intended to be a continually evolving database, which we host online at \href{https://osprei.space/llamacore}{osprei.space/llamacore}. This work presents the first iteration of LLAMACoRE. In this first version, we focus on the time period 2007--2014, which is when both STEREO spacecraft were active and the largest number of 3D CME reconstructions have been performed using white-light data from the C2 and C3 cameras part of the Large Angle and Spectrometric Coronagraph \cite<LASCO;>{Bru95} on board SOHO and/or the COR1 and COR2 telescopes part of the Sun Earth Connection Coronal and Heliospheric Investigation \cite<SECCHI;>{How08} on board STEREO.  We emphasize that no new reconstructions were performed for this work, we have simply compiled existing results and identified matching cases between catalogs. Before analyzing the combined results we present the individual catalogs incorporated into LLAMACoRe.

\subsection{Individual Catalogs}
We use 21 different source catalogs to build LLAMACoRe. Each of these catalogs uses imagery from at least two satellites to reconstruct the 3D direction of a CME at at least one point during its coronal propagation. At a minimum, each catalog contains the reconstructed latitude and longitude for each of their events and contains at least five events. These must be a reconstructed position of the CME apex in the corona and not simply the initial direction of the source region of the CME. We also do not include catalogs that only include the position angle as this only provides a projected location on the plane of sky, not a distinct location in 3D space. A few sources reconstruct CMEs multiple times using different techniques (e.g., both GCS and ICC). Some sources use the positional information from a different source but add in new, additional values, such as the velocity or mass. We have refined these 21 sources into 24 different catalogs, splitting the sources with multiple techniques and combining those with repeated results. 

Table~\ref{cats} lists these catalogs by name, either by their given name, in the case of large established sets, or by their author's last name. It also includes the appropriate reference, the time range covered by the catalog, the number of events (nEvents), the number of events that appear in at least one other catalog (nMulti), and the parameters that the catalog includes. For the parameters, we collect the latitude (lat), longitude (lon), orientation or tilt (tilt), angular width (AW), GCS kappa ($\kappa$), velocity (v), and mass (M). We note that the AW is the half width, the common version typically reported in the literature, and there is some subtlety in comparing the AW from different reconstruction techniques, which we largely ignore. The $\kappa$ parameter is the least intuitive of these parameters, and it is essentially an aspect ratio. For a constant AW, increasing $\kappa$ will make the CME ``thicker'' or ``fatter'' in the edge-on direction. Its mathematical definition can be found in \citeA{The06}. Some catalogs may have additional features, such as CME source regions or associations with in situ events, but we do not include these within LLAMACoRE at this time. If a catalog includes multiple coronal reconstructions for the same event at different distances/times then we only include the outermost one, which should incorporate the most coronal evolution of the CME and correspond to its properties as the CME begins its interplanetary journey. The following list briefly describes each of our 24 catalogs.

\begin{enumerate}
    \item \textbf{AFFECTScat} \cite{Bos12} -- The Advanced Forecast For Ensuring Communications Through Space (AFFECTS) catalog began as a Ph.D.\ thesis project and continued with project support from the 7th Framework Programme of the European Union. AFFECTs focused on the STEREO SECCHI/COR2 field of view and identified 1071 CMEs. A subset of these was then fit with the SWPC-CAT model. 
    \item \textbf{AFFECTSgcs} \cite{Bos12} -- Analogous to the AFFECTScat set but for a subset of cases for which GCS reconstructions were performed. The KinCAT catalog builds upon the AFFECTS GCS reconstructions, using the same position and size, but adds in the velocity and mass. Our ``AFFECTSgcs'' catalog only includes the CMEs that have AFFECTS GCS reconstruction but no additional KinCAT data. 
    \item \textbf{AFFKin} \cite{Bos12,Plu19} -- The overlap between the AFFECTS GCS and the KinCAT catalogs, which we refer to as ``AFFKin.'' Reconstructions that are included in AFFKin are not included in AFFECTSgcs or our ``KinCAT'' catalog.
    \item \textbf{Braga} \cite{Bra17} -- The CORSET3D method is applied to STEREO SECCHI/COR2 images. This method updates CORSET \cite{Gou10}, a supervised computer vision algorithm, to function in 3D using tie-pointing and triangulation. 
    \item \textbf{DONKI} \cite{DONKI} -- The Database Of Notifications, Knowledge, Information (DONKI) is an online repository of space weather information generated by the Community Coordinated Modeling Center (CCMC) at NASA Goddard Space Flight Center. While the CCMC originally generated the DONKI content, after 2020 the CCMC only hosts the repository, and the content is generated by forecasters at the Moon to Mars Space Weather Analysis Office. Of interest to LLAMACoRE, DONKI contains reconstructions (largely performed using the StereoCAT and SWPC-CAT tools) for CMEs observed by the coronagraphs from both STEREO satellites and LASCO. As the content is generated from continuous, daily forecasts, this is the most complete catalog for the time span it covers.
    \item \textbf{Gopalswamy} \cite{Gop14cat} -- This work reconstructs CMEs associated with energetic particle events. The GCS technique is applied to simultaneous STEREO and SOHO coronagraph images. We note that \citeA{Gop14cat} split their catalog into smaller lists based on particle association and source location. Three of their events appear multiple times within the smaller lists and we only include these events once. For two of the three, the reconstructed CME properties are the same across lists. The 2013-05-15T01:25 CME is listed once with the coronal position matching its source region and once with a deflected position. We assume the deflected position within LLAMACoRe.
    \item \textbf{Gui} \cite{Gui11} -- This work analyzes the coronal deflections of a small set of CMEs. They use the GCS technique to reconstruct each CME's position and orientation at multiple coronal distances. For LLAMACoRe, we only include the outermost reconstructions, which should incorporate most, if not all, of the CME's nonradial motion.
    \item \textbf{Isavnin} \cite{Isa13} -- This work compares the coronal reconstructions of CMEs with their interplanetary orientations, as inferred from Grad--Shafranov reconstructions. The CMEs are reconstructed in the corona using the GCS technique with STEREO and LASCO images.
    \item \textbf{Jang} \cite{Jan16} -- This work uses the StereoCAT technique with STEREO images to reconstruct front-sided (from Earth's perspective) halo CMEs that occurred between 2009 and 2013. A major focus of the project was comparing the direction and speed calculated with 3D techniques as opposed to previous 2D techniques. We only include the 3D parameters in LLAMACoRe as they should be more accurate, in general. 
    \item \textbf{Kay} \cite{Kay1745} -- This work includes reconstructions for 45 Earth-directed STEREO-era CMEs and the corresponding expected in situ profiles. The CMEs were reconstructed from STEREO images using the GCS technique.
    \item \textbf{KinCAT} \cite{Plu19} -- The Coronal Mass Ejection Kinematic Database Catalogue (KinCAT) is a part of the Heliospheric Cataloguing, Analysis and Techniques Service \cite<HELCATS,>{HELCATS}. HELCATS largely focuses on CMEs observed in Heliospheric Imagers, but the KinCAT portion helps connect it with the coronal reconstruction of CMEs. The reconstructed positions are mostly the exact same as the AFFECTS catalog, but it includes a few additional cases, as well as CME mass and velocity. The majority of the cases from \citeA{Plu19} are incorporated into our AFFKin catalog, we only label events as KinCAT if they do not also appear in the AFFECTS catalog.
    \item \textbf{Liewer} \cite{Lie11} -- This work compares new reconstructions made with the tie-pointing and triangulation to previously made GCS reconstructions. The work determines the reliability of a forecasted trajectory given the relative locations of the CME and observing satellites. We only include the new reconstructions from \citeA{Lie11} as the GCS reconstructions were originally from \citeA{The09} and are already a part of the Rodriguez catalog.
    \item \textbf{Majumdar} \cite{Maj20} -- This work analyzes the coronal expansion and propagation of a set of CMEs. The GCS technique is applied to STEREO images to reconstruct 3D kinematic profiles. 
    \item \textbf{Martinic} \cite{Mar22} -- This work reconstructs isolated events to compare with the in situ orientation. Events are reconstructed with the GCS technique using STEREO and SOHO coronagraph images. They also reconstruct the orientation of the CME by aligning an ellipse to just a SOHO image, but do not include latitude/longitude measurements with these orientations so we do not include them in LLAMACoRe. Their last event is from 2016 so it is also not included in LLAMACoRe at this time.
    \item \textbf{Rodriguez} \cite{Rod11} -- This work uses coronal reconstructions to test whether one can predict the occurrence of corresponding in situ arrival. The CMEs are reconstructed from STEREO images using the GCS technique. We note that the first 26 of the 34 CMEs within this catalog have reconstructions that were originally published in \citeA{The09}.
    \item \textbf{Sachdeva} \cite{Sac17} -- This work performs an analysis of the drag and Lorentz forces influencing the propagation of a CME. The GCS technique is used with SOHO/LASCO and STEREO images to reconstruct the CME trajectory for comparison with a force-based model.  
    \item \textbf{Shen} \cite{She14b} -- This work reconstructs a series of front-sided CMEs in order to determine whether each event is likely to impact the Earth or not. The GCS technique is applied to reconstruct the CMEs in the STEREO/COR2 field of view.
    \item \textbf{Shi} \cite{Shi15} -- This work uses CME reconstructions to initiate a drag-based model and determine the expected transit time for each event. Events with unambiguous shock fronts are reconstructed by fitting the GCS model to STEREO and SOHO observations.
    \item \textbf{Temmer09} \cite{Tem09} -- A geometric triangulation technique is used to reconstruct CMEs from STEREO and SOHO images. They create profiles of the velocity through the corona as seen from each viewpoint, then use them to infer the source location. We take the average of the position reported for the STEREO-A and STEREO-B views and use their deprojected velocity.
    \item \textbf{Temmer21} \cite{Tem21b} -- This catalog combines STEREO GCS reconstructions from several other catalogs, in addition to providing several new measurements. It includes reconstructions from both the AFFECTS and Majumdar catalogs, so we remove any overlapping cases to make sure the results are not duplicated within LLAMACoRe.
    \item \textbf{Wood} \cite{Woo17} -- This work connects coronal CME observations with in situ observations of magnetic clouds. The CMEs are reconstructed from STEREO coronagraph imagery using the technique from \citeA{Woo09}, which is conceptually similar to visually aligning the GCS shape, but employs a different set of equations defining the shape.
    \item \textbf{Zhong} \cite{Zho21} -- This work uses the GCS technique to reconstruct Earth-impacting CMEs from the corresponding STEREO coronagraph observations. Their fits include several reconstructions from the Shen catalog, which we do not include as duplicates within LLAMACoRe.
    \item \textbf{ZhuangGCS} \cite{Zhu17} -- This work uses coronal reconstructions of CMEs as an input to the Integrated CME-Arrival forecasting (iCAF) system. There are both GCS and ICC reconstructions, which we separate into two different LLAMACoRE catalogs for comparison. This is the GCS component of the Zhuang catalog.
    \item \textbf{ZhuangICC} \cite{Zhu17} -- This is the ICC component of the Zhuang catalog.
\end{enumerate}

\begin{sidewaystable}
    \centering
        \caption{List of the individual catalogs within LLAMACoRe}
    \begin{tabular}{rcccccl}
          & Name & Reference & Time Range & nEvents & nMulti & Parameters \\
         \hline
       1  & AFFECTScat & \citeA{Bos12} & May 2007 -- Dec 2011 & 196 & 196 & Lat, Lon, AW, v \\
       2  & AFFECTSgcs & \citeA{Bos12} & Jan 2007 -- Dec 2011 & 139 & 120 & Lat, Lon, Tilt, AW, k \\
       3  & AFFKIN & combines 2 and 10 & May 2007 -- Dec 2011 & 102 & 97 & Lat, Lon, Tilt, AW, k, v, M \\
       4  & Braga & \citeA{Bra17} & Dec 2008 -- Nov 2011 &  17 & 17 & Lat, Lon, v \\
       5  & DONKI & \citeA{DONKI}  & Apr 2010 -- Dec 2014 & 1611 & 380 & Lat, Lon, AW, v \\
       6  & Gopalswamy & \citeA{Gop14cat} & Feb 2010 -- Jan 2014 & 74 & 65 & Lat, Lon, v\\
       7  & Gui & \citeA{Gui11} & Nov 2007 -- Dec 2008 & 10 & 10 & Lat, Lon, Tilt, AW, k, v\\
       8  & Isavnin & \citeA{Isa13} & Jun 2008 -- Dec 2010 & 15 & 11 & Lat, Lon, Tilt \\
       9  & Jang & \citeA{Jan16} & Oct 2009 -- Dec 2013 & 306 & 282 & Lat, Lon, AW, v \\
       10  & Kay & \citeA{Kay1745} & Nov 2007 -- Jun 2014 & 45 & 44 & Lat, Lon, Tilt, AW, k, v, M \\
       11  & KinCAT & \citeA{Plu19} & Dec 2007 -- Oct 2013 & 20 &  20 & Lat, Lon, Tilt, AW, k, v, M\\
       12  & Liewer & \citeA{Lie11} & Aug 2007 -- Apr 2008 & 7 &  6 & Lat, Lon, v \\
       13  & Majumdar & \citeA{Maj20} & May 2007 -- Apr 2014 & 59 & 45 & Lat, Lon, Tilt, AW, k, v \\
       14  & Martinic & \citeA{Mar22} & Dec 2008 -- Aug 2014 & 21 & 19 & Lat, Lon, Tilt, AW, k \\
       15  & Rodriguez & \citeA{Rod11} & Nov 2007 -- Dec 2008 & 34 & 23 & Lat, Lon \\
       16  & Sachdeva & \citeA{Sac17} & Mar 2010 -- Dec 2013 & 38 & 37 & Lat, Lon, Tilt, AW, k, v\\
       17  & Shen & \citeA{She14b} & Dec 2009 -- May 2012 &  39 & 39 & Lat, Lon, AW, v\\
       18  & Shi & \citeA{Shi15} & Dec 2008 -- Oct 2012 &  21 & 21 & Lat, Lon, AW, v\\
       19  & Temmer09 & \citeA{Tem09} & May 2007 -- Mar 2008 & 10 & 9 & Lat, Lon, Tilt, AW, k, v, M \\
       20  & Temmer21 & \citeA{Tem21b} & Apr 2010 -- Aug 2014 & 22 & 22 & Lat, Lon, Tilt, AW, k, v, M \\
       21  & Wood & \citeA{Woo17} & Dec 2008 -- Jul 2012 & 28 & 18 & Lat, Lon, Tilt, AW, v \\
       22  & Zhong & \citeA{Zho21} & Jul 2008 -- Dec 2012 & 71 & 52 & Lat, Lon, Tilt, AW, v\\
       23  & ZhuangGCS & \citeA{Zhu17} & Feb 2010 -- Apr 2012 & 31 & 31 & Lat, Lon, AW, v \\
       24  & ZhuangICC & \citeA{Zhu17} & Feb 2010 -- Apr 2012 & 38 & 38 & Lat, Lon, AW, v \\
    \end{tabular}
    \label{cats}
\end{sidewaystable}

We note that LLAMACoRE presents the CME longitude exclusively in Stonyhurst coordinates. The source catalogs use a mix of Stonyhurst and Carrington coordinates for the reconstructions. For those in Carrington coordinates, we determine the corresponding Carrington longitude of Earth at the given time of the reconstruction and use this to convert the CME longitude into Stonyhurst coordinates. In other words, all LLAMACoRe positions are reported within a coordinate system such that 0$^{\circ}$ latitude and longitude corresponds to an Earth-directed CME.

\subsection{Connecting Reconstructions to Single Events}
We start assembling LLAMACoRe by defining a master list of all CMEs that occurred between 2007--2014. AFFECTS and DONKI are by far the largest catalogs so we first combine their CMEs into a list of unique CMEs. Each CME is identified by a unique timestamp corresponding to some reconstruction time during its coronal propagation, taken from one of the source catalogs. There is only a small window of overlap between these two catalogs, between April 2010 and December 2011, so we use all the AFFECTS events (and their associated times) before April 2010 and all the DONKI events after December 2011

We must stitch together the two catalogs in the overlapping time range. We start by adding the full list of all AFFECTS CMEs over this time to the master list. We begin with AFFECTS as the DONKI catalog appears to be less thorough during the earliest days of their routine measurements.  We take each DONKI event between April 2010 and December 2011 and determine if there is a close time match in the AFFECTS catalog. For any single event, the reconstruction time is probably not the exact same between different catalogs, potentially varying by a few hours depending on which images were used. If the distance of reconstruction is given for both we can check if the time differences make sense (i.e.\ the farther reconstruction \textit{should} correspond to a later time). 

We then compare the reconstructed geometry to see if we can validate the plausible match. We primarily compare the latitude and longitude as we have those parameters for every reconstruction. We expect there to be some uncertainty in these values, but we check whether the reconstructions fall in the same general area. If the positions fall in completely different hemispheres it is fairly unlikely, though not completely impossible, that they are reconstructions of the same event. If the times and positions seem reasonable then we associate the DONKI reconstruction with the timestamp ID of the corresponding AFFECTS event. For the majority of reconstructions, this is a clear decision to associate or not, but there are a few events where the decision is somewhat subjective. In general, we tend to make the conservative decision and not associate events unless we are certain.

We take the master list of times and add in any DONKI events that were not already included during the overlapping time period. We then repeat the process for each other catalog, associating each event with a corresponding existing case or adding the CME if we cannot find a suitable match. At the end of the process, we have a master list with a unique timestamp ID for every CME and every reconstruction in every catalog is associated with one of these IDs. We initiated our master list with the full list of all AFFECTS CMEs, which contains the times of all CMEs over its range, not just those with AFFECTS GCS or CAT reconstructions. We now remove any of those times that do not have a corresponding reconstruction, either from AFFECTS itself or some other catalog, so that our list only contains entries with reconstructions. This leaves us with a master list of 1862 CMEs with at least one reconstruction and a set of 2954 reconstructions across all the catalogs. We find that for 511 CMEs we have more than one reconstruction, which we will refer to as ``multi-cat'' events. The full list of all reconstructions for all CMEs is included in the supplementary material of this manuscript. This corresponds to the version of LLAMACoRe used in this work, updates will be made to the online version as additional reconstructions become available. We also include a text file with one best set of parameters per CME, as well as their uncertainties (when available). The details of deriving these best parameters are presented in Section~\ref{gen} and the corresponding uncertainties in Section~\ref{multiSec}.

\begin{figure}[ht!]
\includegraphics[width=\textwidth]{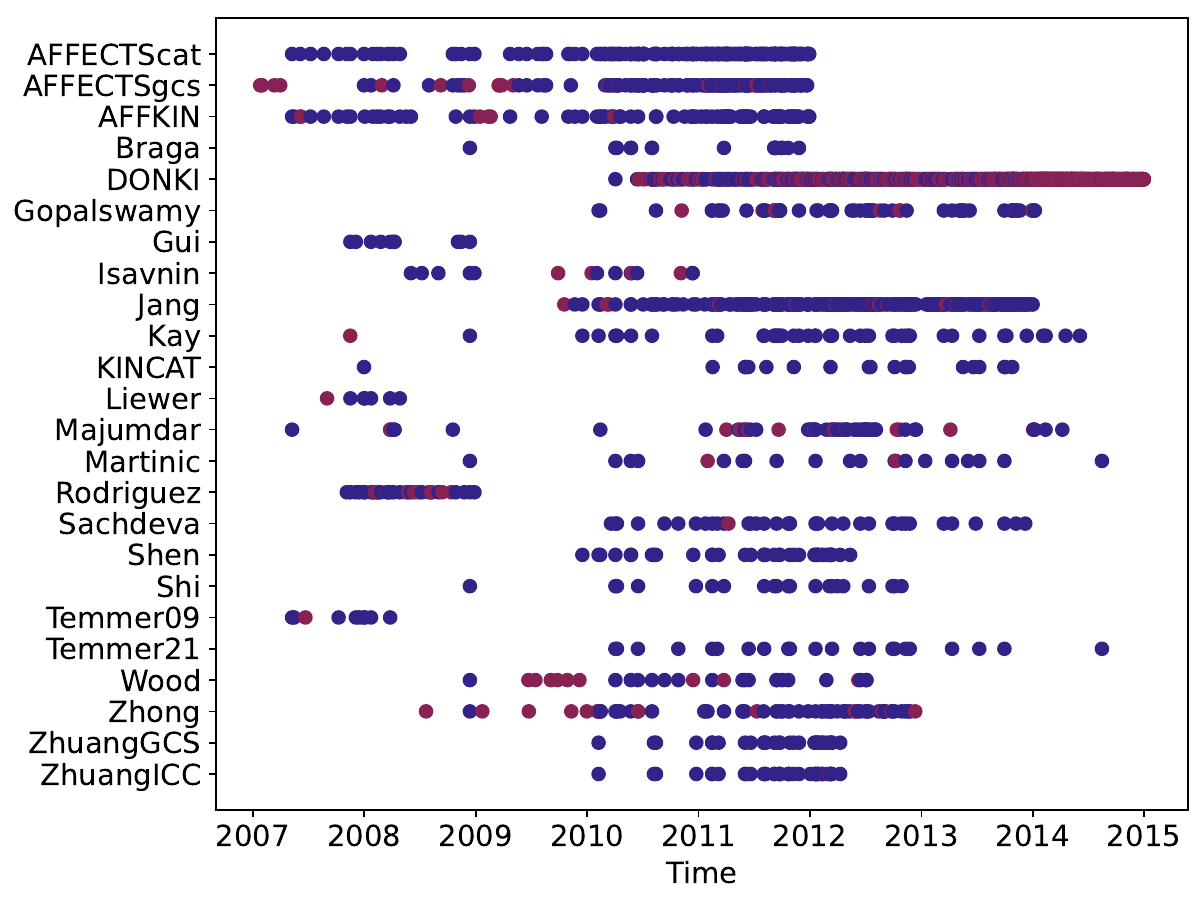}
\caption{Temporal span of the individual catalogs used within LLAMACoRe. Each dot represents a reconstructed CME, the maroon points correspond to events that only appear in a single catalog whereas the blue dots are multi-catalog events.} \label{span}
\end{figure}

Figure~\ref{span} shows the spread of the events within each catalog. The individual catalogs are listed vertically and the horizontal axis shows time. Each dot represents the time of the individual reconstructions within a catalog. The maroon dots correspond to CMEs with only a single reconstructions and the blue dots are the multi-cat events. This figure shows the range of catalogs included within LLAMACoRE, which varies from large catalogs covering nearly all events over a long time frame to smaller catalogs with only a few case studies.

The supporting information includes a table with the number of overlapping cases between each pair of catalogs. For each catalog, it also shows the total number of reconstructions (nRecon) and the number of reconstructions that have no overlap with any other catalog. AFFECTS and DONKI are the most extensive catalogs but have minimal temporal overlap between the two, so we expect them to have a high number of no-overlap cases. In the other catalogs, we double-check the no-overlap events to ensure we have not missed any plausible connections. This corresponds to a total of 98 no-overlap events. All of these do not readily match any other entry with the closest event to each having either a difference of $>$12 hours in time and/or a completely different spatial location. In some cases, the time will be ``close'' to a multi-cat time in the master list, but that catalog has an additional entry that is a better fit both time-wise and geometry-wise. In other cases, there are multiple reconstructions at the same time with differing geometries, and coronagraph observations confirm there are simultaneously multiple distinct CMEs. 

One point we want to acknowledge is that, in addition to the uncertainty in human interpretation of reconstructions, we are likely subject to human errors such as typographical errors when transcribing lists. In developing LLAMACoRE we have attempted to directly copy and paste as much as possible, rather than transcribing by hand, to avoid introducing any additional errors. We suspect that the source catalogs contain some transcription errors, such as adding/skipping a number or forgetting a negative sign, as we have found questionable spread within some events. We first use a Zhuang result as an example, not implying that the Zhaung catalog is any more or less accurate than any other, but it is easy to find examples since they provide both GCS and ICC reconstructions that they themselves have associated with the same event. For example, the 2011-06-21T03:08:15 CME has a GCS longitude of $-20^{\circ}$ but an ICC longitude $80^{\circ}$. We expect there to be some variation between different techniques, but these are essentially perpendicular directions. As another example, the 2012-03-12T02:42:00 CME has reconstructed latitudes of $35^{\circ}$, $-37^{\circ}$, and $-53^{\circ}$ in the DONKI, Jang, and Majumdar catalogs, which is the largest spread in latitude for a single event within LLAMACoRe. We have confirmed that DONKI lists the latitude as positive, but recognize that if they happened to have forgotten a negative sign it would make much better agreement between the different reconstructions. In any similar situations, we have verified that we are correctly reproducing the published results from each catalog, but we cannot confirm whether the large spreads are real features of using different reconstruction techniques or less scientifically relevant human errors. We note that, at some level, all the results presented here are a combination of these errors. We also point out that such errors could explain some of the no-overlap events. We will revise LLAMACoRe if we receive confirmation of human errors that should be corrected.

\section{General CME Properties}\label{gen}
After collecting all the events for the meta-catalog, we present an analysis of the reconstructions within this first version of LLAMACoRE. We show a summary of the general CME properties within the set in this section and an analysis of the uncertainty and spread in the reconstructed CME properties using the multi-cat events in Section \ref{multiSec}.

\subsection{Distribution of Properties}
Since we have multiple reconstructions of many of the CMEs, we can combine them into what is hopefully the most accurate reconstruction of a CME's properties. We have no relative judgment of the individual catalogs or reasons to favor one over the other.  Assuming all catalogs are of equal quality, then combining them may average out some of the uncertainties due to the subjective nature of CME reconstructions. For each CME, we determine the median value of the reconstructed properties using whatever information is available from the combined catalogs. For the single-cat cases, this is the single reconstruction. For the multi-cat cases, this is the median for each parameter, and a different number of catalogs may be available for different parameters. For example, all catalogs include latitude and longitude, but very few have masses, so the median properties may include multi-cat measurements of the position but a single-cat measurement of the mass. We note that circular medians were used for the longitude and tilt. We use the median instead of the mean as it is less sensitive to outliers. We will refer to this set of median properties as the ``best-constrained'' values throughout this work.

\begin{figure}[ht!]
\includegraphics[width=\textwidth]{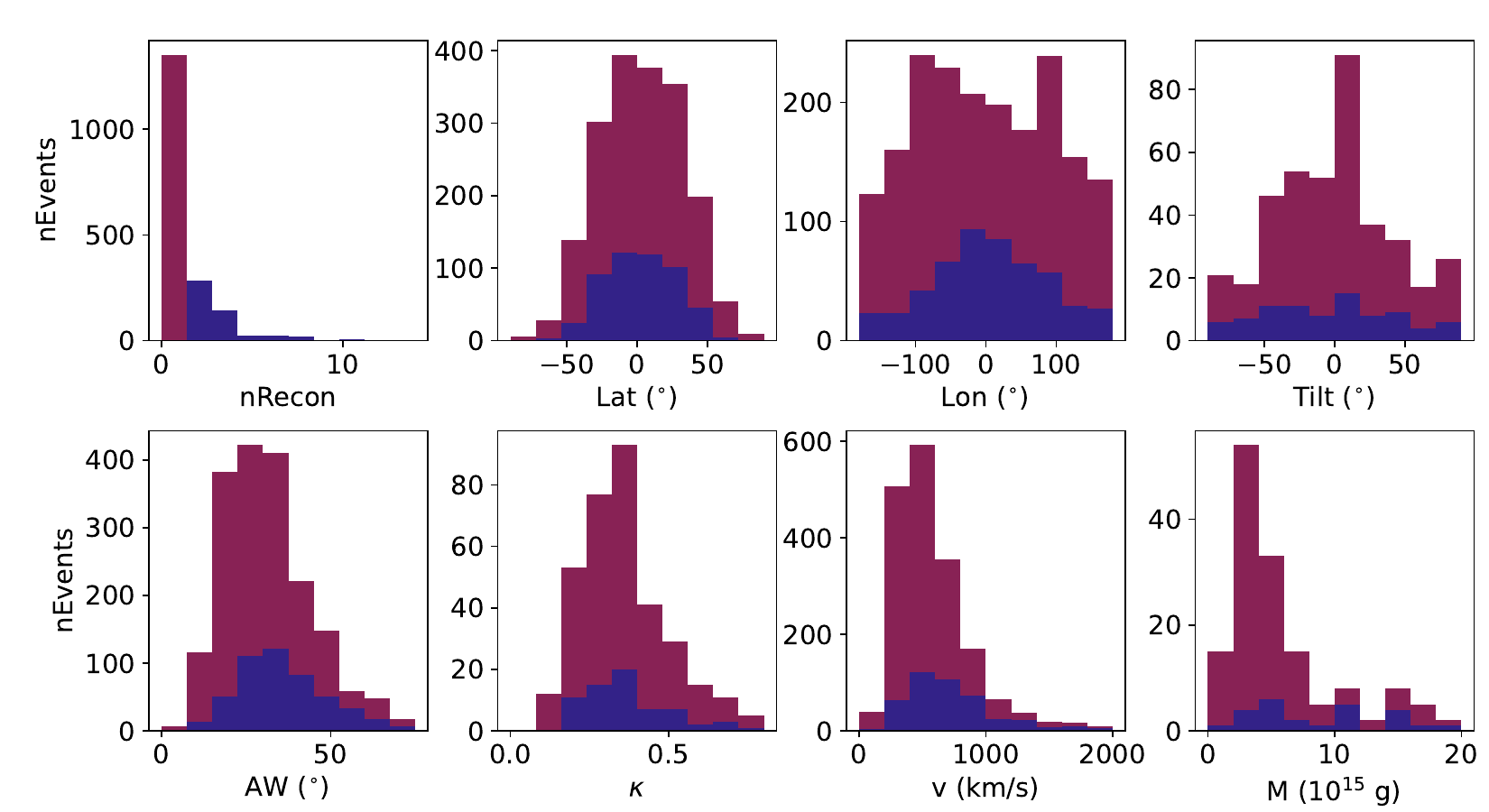}
\caption{Histograms of the best-constrained CME properties. The maroon histograms show the full catalog, including single-source cases, whereas the blue only shows the median values for the multi-cat cases.} \label{histomeds}
\end{figure}

Figure~\ref{histomeds} shows histograms of the best-constrained properties of the LLAMACoRe CMEs. These histograms include each CME once, rather than simultaneously showing all of the reconstructions available for every CME and weighting the distributions by the popular CMEs. If the best-constrained values are the most accurate characterization of CMEs between 2007 and 2014 then it is useful to explore the distribution of observed CME properties. We use the same color scheme as in Figure~\ref{span} with maroon representing the full sample and blue representing only the multi-cat cases. The top row shows the number of reconstructions for each CME (nRecon), the latitude (lat), longitude (lon), and tilt from left to right. The bottom row shows the angular width (AW), GCS aspect ratio ($\kappa$), velocity (v), and mass (M). Note the difference in scale on the y-axis as there are very differing numbers of CMEs for each parameter.

The top left panel shows the large number of single-cat cases compared to multi-cat cases, largely due to the continuous record of nearly all events in DONKI starting in April 2010. We note a slight difference between the two populations depicted in the histograms. This is likely due to a bias toward which events tend to be more popular and have multiple reconstructions versus those that receive less attention and have been reconstructed only once. This bias is clearly reflected in the histograms. For example, the multiple reconstruction events tend to have smaller absolute longitudes and, to a lesser extent, smaller absolute latitudes. We can also see that the multi-cat events are disproportionately missing more of the smaller AW and slower CMEs. These trends make sense as the smaller catalogs tend to be biased towards Earth-directed and larger events that are easier to distinguish from the background corona when performing the reconstruction.

Table~\ref{stats} shows statistics for the reconstructed CME properties, providing estimates of the most probable CME values and their ranges. We include measures for both the full set of reconstructions (labeled ``All'') and the multi-cat set (labeled ``Multi''). We show the number of events (n), the median (Med) of the best-constrained values, and the interquartile range (IQR). The IQR is the range between the 25th and 75th percentiles and shows the spread of the middle half of the distribution. For the longitude and tilt, which are circular distributions, we sort the values and then shift them one by one across the circular boundary (e.g.\ moving a longitude of $-170^{\circ}$ to $190^{\circ}$) to determine where the distribution extends over the shortest range. We then determine the median and IQR of this distribution.

\begin{table}[ht!]
    \centering
    \caption{Number of events, medians (Meds) and interquartile range (IQR) of the full set of CMEs and the multi-cat CMEs.}
    \label{stats}
    \begin{tabular}{r|r r r | r r r }
        Property & n$_\mathrm{All}$ & Med$_\mathrm{All}$ & IQR$_\mathrm{All}$ & n$_\mathrm{Multi}$ &  Med$_\mathrm{Multi}$ & IQR$_\mathrm{Multi}$  \\
        \hline
       Lat ($^{\circ}$) & 1862 & 2.8 & 44.6 & 511  & 2.6 & 38.5  \\
       Lon ($^{\circ}$) & 1862 & -5.0 & 161.0 & 511 & 1.6 &  112.7 \\
       Tilt ($^{\circ}$) & 394 & 0 & 55.0  & 320  &  -0.5 & 58.2 \\
       AW ($^{\circ}$) & 1833 & 29.0 & 17.0 & 508 & 33.3 & 18.3 \\
       $\kappa$ & 341 & 0.34  & 0.17 & 300  & 0.35 & 0.18 \\
       v (km/s)  & 1822 & 509  & 345 &  507 & 619 &  428 \\
       M (10$^{15}$ g)  & 157 & 4.6 & 5.2 & 151  & 4.6  & 5.6
   \end{tabular}
\end{table}

These numbers confirm the slight bias in the multi-cat distribution, as the median speed and AW are slightly larger than those of the full distribution, highlighting the trend for more people to fit more energetic or dynamic CMEs. If we want to analyze how ``Earth-directed'' the longitude is, it is best to look at the median value of the absolute longitude (not shown in the Table), rather than the median of the circular range (shown in the Table). For the median absolute longitude, we find $57.0^{\circ}$ for the multi-cat events as opposed to $80.5^{\circ}$ for the full set, hinting more strongly at the popularity of Earth-directed events in the multi-cat set.

In general, the median properties of the full set are not the most useful as there is a wide variance in them throughout the solar cycle, as evidenced by the IQR for both the full set and the multi-cat set. We do find a median CME velocity of 509~km/s, which is slightly higher than the generally quoted ${\sim}400$--450~km/s \cite<e.g.,>{Com17, Yur05}, but again this is likely simply a bias of which CMEs tend to be more suitable for 3D reconstructions. Reconstructions are more likely to be performed for brighter CMEs that are easy to identify in coronagraph images, which tend to be in turn more energetic---the CME kinetic energy depending on mass and speed \cite<e.g.,>{Vou10}. In fact, the median CME velocity for the multi-cat cases is 619~km/s, proving again that case studies tend to focus on faster and generally more energetic CMEs.

\subsection{Temporal Variations}
\begin{figure}[ht!]
\includegraphics[width=\textwidth]{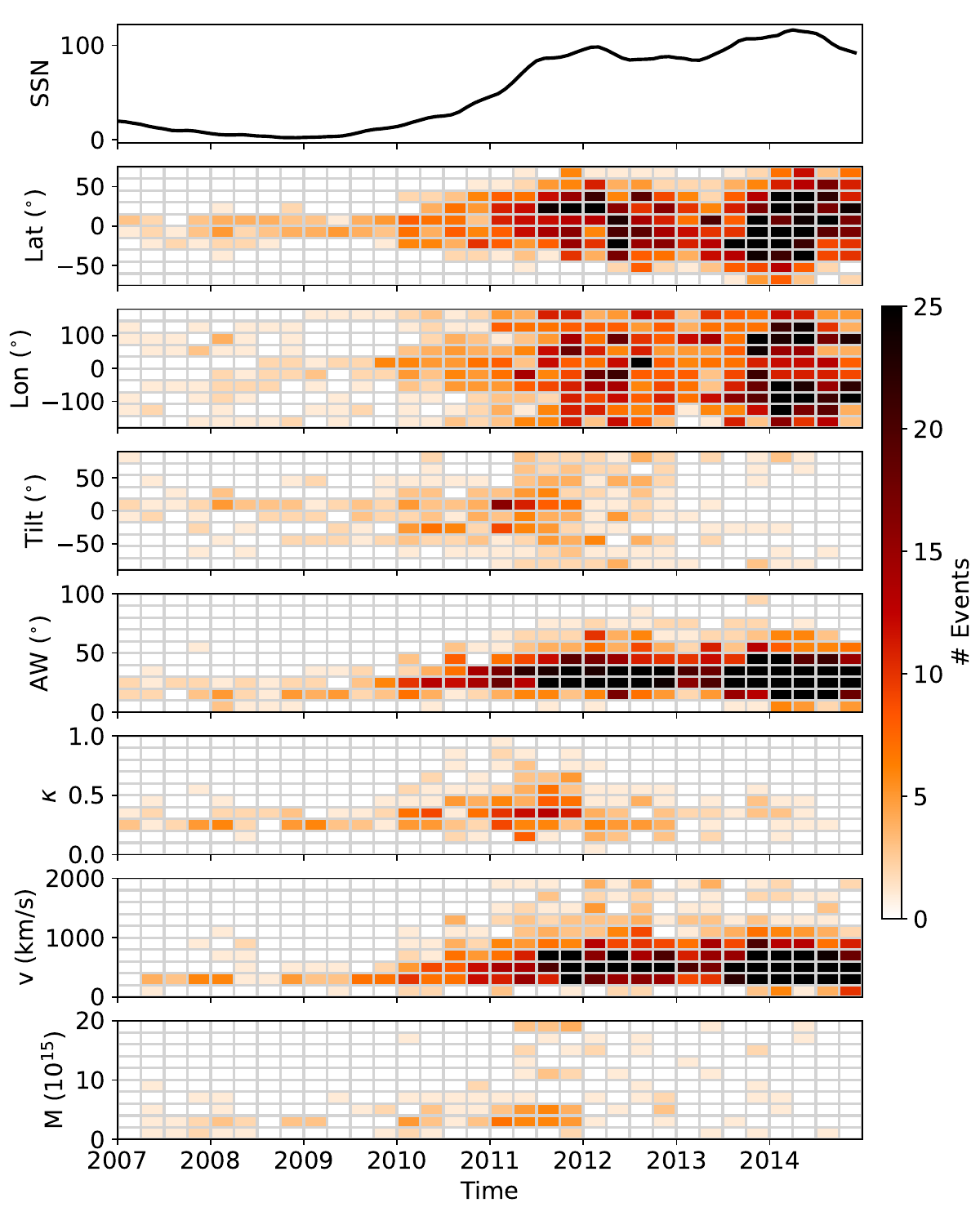}
\caption{Heat map showing the range of reconstruction CME properties as a function of time. The grid cell color indicates the number of events with that given property value at a given time. The top row shows the smoothed sunspot number.} \label{timelinemedians}
\end{figure}

To investigate the solar cycle dependence we look at the spread in CME properties as a function of time. Again, we are still considering the properties of individual CMEs, whether that is a single reconstruction or the median for a multi-cat case. Figure \ref{timelinemedians} shows a heat map where we have translated both time and CME properties onto a uniform grid and tallied the number of events that fall within each grid cell. Darker cells represent more counts and white indicates no instance of that property value at that time. To facilitate comparison to the solar cycle, the top row of Fig. \ref{timelinemedians} shows the smoothed sunspot number (SSN) using data retrieved from \citeA{sidc}. 

We see a general increase in the number of reconstructed CMEs over our considered time range, which is a combination of several factors. First, there is the solar cycle dependence with a greater number of events occurring as the Sun approaches solar maximum, which occurred in April 2014 for Solar Cycle 24. Second, there is the feasibility of performing multi-viewpoint reconstructions. At the launch of STEREO, there was minimal angular separation between the two satellites. They began slowly separating, at some point reaching an optimal separation for stereoscopic reconstruction of Earth-directed CMEs, then continuing onward. By 2011 they were located around $\pm$90$^{\circ}$ from Earth and by the end of 2014 they reached the far side of the Sun and communication was lost with STEREO-B, making stereoscopic reconstructions more difficult.

We see some trend in the reconstructed latitude as a function of the solar cycle. Toward solar minimum, the CMEs tend to be located closer to the equator, but they are spread over a larger range of latitudes as activity increases. This results as a combination of the source regions of CME and their coronal deflection \cite<e.g.,>{Cre04, Gop09}. The most energetic CMEs tend to erupt from active regions, which, at the start of the rise phase of a new solar cycle (about 2010 here), tend to first appear toward higher latitudes, then tend toward emerging at lower latitudes as the cycle progresses. Deflection, however, tends to cause CMEs to move away from coronal holes and toward the heliospheric current sheet (HCS), following the gradients in the background magnetic energy \cite<e.g.,>{Cre06, Kay15}. At solar minimum, the HCS has low inclination, and we see more CMEs near the equator than at high latitudes. As the solar cycle progresses, the active regions move to lower latitudes but the HCS becomes more inclined, and we see a wider range of reconstructed latitudes. We see low latitude reconstructions throughout the end of Solar Cycle 23, the minimum between cycles, and still in much of the rise phase of Solar Cycle 24. As maximum approaches, the reconstructions exhibit a wider range of latitudes.

We do not note any strong trends in the longitude as a function of time but one would not expect there to be a preferred longitude for CMEs.  We also cannot identify any trend in tilt, $\kappa$, or $M$ over time, but this is strongly affected by the limited number of events with reconstructions of these parameters. We see larger speeds and AWs as solar maximum approaches. There is a trend toward slightly higher values for the highest density grid cells (the core of the distribution), but we also encounter a larger number of outlier events with ``extreme'' CME speeds as solar activity increases. Overall, we note nothing fundamentally new in these results but we can confirm that the combined LLAMACoRe catalog reproduces the generally expected behavior of CMEs \cite<e.g.>{Gop20, Lam19, Yas04}.

\subsection{Correlation Between Parameters}
Our final analysis of the best-constrained properties, before looking at the scatter within the multi-cat cases, is looking for any relations between reconstructed CME properties. The literature contains many examples attempting to relate parameters such as angular width or mass to velocity \cite<e.g.,>{Plu19, Vrs05, Vrs08, Vou02, Wel18}. Such relations, especially those relating a parameter that is difficult to measure with one that is not (such as relating mass to velocity) could be extremely useful for any space weather predictions that require forward-modeling a CME starting in the outer corona.

\begin{figure}[ht!]
\includegraphics[width=\textwidth]{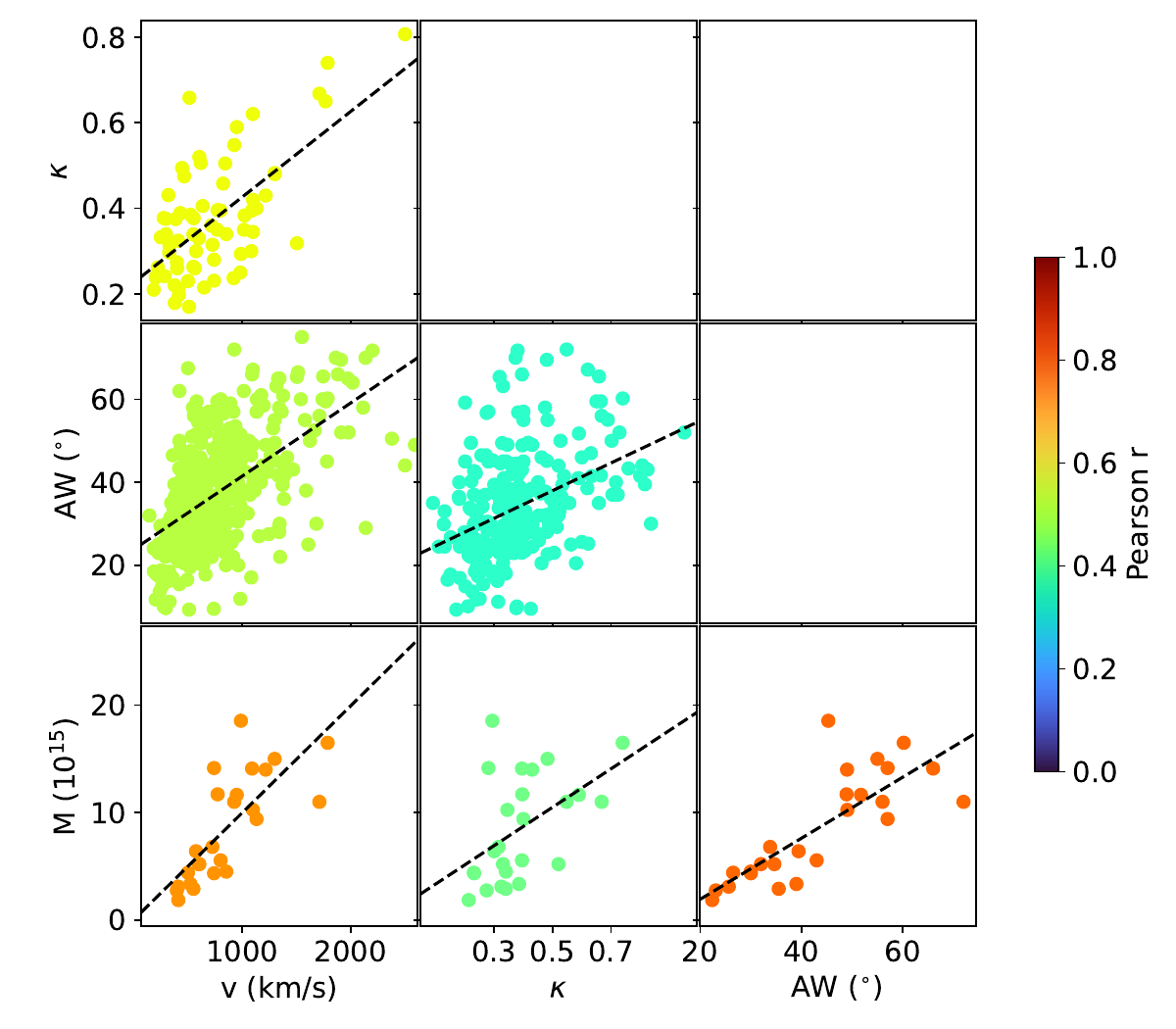}
\caption{Correlations between some of the best-constrained properties for the multi-cat cases. The panels are colored by the Pearson r-value and the black dashed line shows the linear regression between those two properties.}\label{corrmeds}
\end{figure}

Figure~\ref{corrmeds} shows correlations between AW, $\kappa$, v, and M for the multi-cat observations. We use the multi-cat median values, and only specifically the values with multiple reconstructions, as we hope they are more accurate than the individual reconstructions, which should lead to stronger correlations. For example, while there are 151 events in the multi-cat sample with a measured mass, which are considered multi-cat because multiple catalogs include at least their latitude and longitude, only 25 of those had the mass reconstructed in two or more sources. We have multiple reconstructions for the AW for 485 events, of $\kappa$ for 67 events, and of v for 507 events. The figure contains a panel with a scatter plot for each pair of parameters. The color of the dots indicates the Pearson correlation coefficient (r-value) for that pair of parameters. The dashed black line shows the linear regression for each set. 

In general, a higher correlation coefficient represents a greater likelihood of a relation between the two parameters, but the value that is ``significant'' depends on the size of the samples being compared. The correlation coefficient and sample size can be used to compute the probability that the correlation could accidentally appear from random uncorrelated distributions.  Using a somewhat arbitrary cutoff of a less than 1\% chance of being an artificial correlation, we find significant correlations between all pairs except for $\kappa$ and M. We list the linear regressions for the significant correlations below.
\begin{eqnarray}
    \mathrm{AW} &=& 0.0176\; \mathrm{v} + 23.8 \\
    \kappa &=& 0.000200\; \mathrm{v} + 0.226 \\
    \kappa &=& 0.00495\; \mathrm{AW} + 0.210 \\
    \mathrm{M} &=& 0.010\; \mathrm{v} + 0.016 \\
    \mathrm{M} &=& 0.284\; \mathrm{AW} - 3.76
\end{eqnarray}
where v is always in km/s, AW in $^{\circ}$, and M is in 10$^{15}$ g. Note that we have inverted the equation relating AW and $\kappa$ from the figure to give what we believe is the more useful form. These regressions are certainly biased based on the multi-cat cases tending to be larger, faster cases. They imply an AW of 22.6$^{\circ}$ and an M of 1.6$\times$10$^{14}$ g for a zero-velocity CME. There are plenty of observations of CMEs smaller than the AW limit, so we suggest caution in blindly applying these relations but we provide them as a useful characterization given the reconstructions we currently have.

We do not include the latitude, longitude, and tilt in the pairings shown in Figure~\ref{corrmeds}. To first order, we do not expect these parameters to scale with the CME intensity (e.g., if CMEs from the west tended to be faster). One could make an argument, however, that solar cycle variations might simultaneously affect the intensity and the latitude and tilt through the variations in the source region, deflection, and rotation. We checked for such variations but found no significant correlations. We also note that we have repeated this analysis using the full set of CMEs but the correlations are much weaker than what we find for using exclusively the multi-cat events. We propose that the multi-cat regressions are the optimal versions to use, but have not tested anything against additional observed events.

\section{Variation Within Multi-Cat Events}\label{multiSec}
We now analyze the scatter between the reconstructions within each multi-cat event. We have already determined the median properties for each case so now we will determine the spread of the reconstructions about those median values. In an ideal scenario, we would determine the quartile values, but the vast majority of our cases only have two or three reconstructions, so we cannot accurately infer that level of detail. Instead, we determine the median absolute deviation (MAD). We determine the absolute difference between each reconstruction and the median value, then take the median of these differences, which is the MAD. This metric is the most appropriate given our sample sizes and the potential for outlier reconstructions. The uncertainties in the supplementary table of the best CME parameters correspond to the MAD. We then take the median value of the set of MADs for all the full multi-cat CMEs, which we refer to as the medMAD. We also determine the IQR of this set of MADs (IQRMAD) and the maximum value (maxMAD). Table~\ref{statsSpread} lists these values for each property, along with the number of events with multiple reconstructions of that property, n$_{\mathrm{MAD}}$.

\begin{table}[ht!]
    \centering
        \caption{Statistics for the spread within the multi-cat cases. We show the number of events with multiple reconstructions of that property, the median MAD over all of the CMEs, the IQR, and the maximum MAD within our sample. We also show the expected deviation between multiple reconstructions of the same event, both in physical units and as a percentage (where appropriate). We also include the mean absolute error in the reconstructed values of synthetic CMEs from \protect\citeA{Ver23}, the average error in the reconstructed values of real CMEs from \protect\citeA{Plu19}, and the uncertainties derived from the sensitivity analysis of the GCS model by \protect\citeA{The09}}.
    \hspace*{-.2in}
    \begin{tabular}{| r | r | r r r | r r | r r r | }
        Property & n$_{\mathrm{MAD}}$ & medMAD & IQRMAD & maxMAD & $\epsilon$ & $\epsilon$ (\%) & V23 & P19 & T09\\
        \hline
       Lat ($^{\circ}$) & 511 & 2.0 & 2.6 & 21.2  & 4.0 & ---  & 1.5 & 5 & 1.8\\
       Lon ($^{\circ}$) & 511 & 4.0 & 5.8 & 50.1  & 8.0 & --- & 3.4 & 5 & 4.3 \\
       Tilt ($^{\circ}$) & 85 & 12.0 & 15.7 & 48.0 & 24.0 & ---  & 2.6 & 30 & 22 \\
       AW ($^{\circ}$) & 485 & 4.7 & 5.8 & 23.0   & 9.3 &  27 & 10.6 & 10 & 10 \\
       $\kappa$ & 67 & 0.05 & 0.06  & 0.28   & 0.1 & 29  & 0.05 & 0.025 & 0.06\\
       v (km/s)  & 448 & 58 & 95  & 837   & 115 & 19 &  --- & --- & --- \\
       M (10$^{15}$ g) & 25 & 1.3 & 2.8 & 6.5  & 2.5 &  38 & --- & --- & --- 
   \end{tabular}
     \label{statsSpread}
\end{table}

These values should be the best-to-date characterization of the uncertainty in CME reconstructions, but the precise statistical definitions are slightly complicated for several reasons. First, we do not have ``true'' values so we can calculate uncertainties but not actual errors. Even calculating uncertainties requires a reference value from which we can calculate the deviation for each reconstruction. We use the median value as the reference point, which helps account for outliers, but makes mathematical interpretation of the uncertainty difficult as the median is not determined from common formulae such as a sum. If we consider a simple example with just two reconstructions, the median value will be halfway in between the two reconstructed values. Each reconstruction will have the same absolute deviation from the median value, so the MAD is the same as the individual deviations in this two-reconstruction example. The difference between the two reconstructions is twice the MAD. 

A similar, general analysis cannot be performed for a generic set of three or more events. The absolute distances between the median and individual reconstructions will be grouped around the MAD, but we cannot say the exact distance between individual reconstructions as it depends on their exact distribution. We suggest that the ``typical'' deviation between two reconstructions of the same event is roughly twice the MAD, but this is an order of magnitude estimate and not a precise calculation. We call this expected deviation $\epsilon$ and propose it is the best estimate of the uncertainty between reconstructed values. In other words, if we have an existing reconstruction and a second, independent reconstruction is performed, we expect the new values to differ by about $\epsilon$. Again this says nothing about the relation of either reconstruction to the ``true'' CME properties, just their relationship to one another. Table~\ref{statsSpread} lists $\epsilon$ for each reconstructed property, both in physical units and as a percentage. The percentages were calculated using the median properties of each event, then reduced to a single value by taking the median of the individual percentages over all of the events.

We compare our uncertainties with the errors found in \citeA{Ver23}, \citeA{Plu19}, and \citeA{The09}, which we will refer to as V23, P19, and T09, respectively. V23 determined the errors in GCS reconstructions using synthetic white-light events, which have known values and for which a perfect reconstruction is possible. This yields true errors for the V23 statistics, rather than just uncertainties. The final column of Table~\ref{statsSpread} lists the mean absolute error (MAE) from V23. We note that we use their values corresponding to all satellite configurations. If we treat the LLAMACoRe medians as true values our medMAD numbers are analogous to the V23 MAE, though there is some subtlety in using a median within LLAMACoRe and a mean within V23. P19 estimated uncertainties in GCS reconstructions based on two independent fits of 15 sample real events, differing from V23 in that the ``true'' CME parameters, in this case, are not known. T09 developed an automated fitting procedure for the GCS model and determined the magnitude of variation that produced a 10\% decrease in their merit function. While this is not the exact same as the variation between visual fits between different observers, it is an interesting comparison. For the T09 errors we show the mean values and for AW and $\kappa$ we show the average of their positive and negative values. We note that all these numbers tend to display similar magnitudes to what is seen in LLAMACoRe with the exception of the tilt, which is significantly lower in V23. This is not surprising, since the axis orientation of a CME is more straightforward to determine for a synthetic, croissant-like structure than for a real, more irregular event.

We see nearly identical metrics for the latitude, longitude, and $\kappa$ between LLAMACoRe and V23. This supports using these values as accurate measures of the uncertainty and potentially the error from the true value. It also builds confidence in our ability to reconstruct these properties for real events since the uncertainties are comparable to the best-case scenario of idealized events. In contrast, we find much larger uncertainty in the tilt but smaller uncertainty in the AW for LLAMACoRe as compared to V23. This is likely a result of comparing the GCS shape to non-idealized, real CMEs as opposed to perfectly symmetrical synthetic CMEs. The difference in the tilt uncertainty follows logically as the synthetic CMEs tend to look more like a distinct, isolated tube with an apparent inclination, whereas real CMEs look more like amorphous blobs that can be hard to distinguish from the background. In terms of the AW, the full extent of the GCS wireframe is a combination of both kappa and AW, with the AW setting the angle between the legs and $\kappa$ determining the width of the cross section. V23 found very large ranges in AW when only a single viewpoint is used, particularly if the event is propagating along the line-of-sight direction. We suspect that for real events most observers tend to minimally vary kappa from the default settings and primarily adjust the AW until the wireframe matches the white-light extent. This is just one potential interpretation, however. We also note that the LLAMACoRe AWs include both toroidal GCS-like shapes as well as more symmetric ICC-like shapes, so it is somewhat surprising to see smaller variations in the LLAMACoRe AWs than in V23.

\subsection{Range in Deviations}
\begin{figure}[ht!]
\includegraphics[width=\textwidth]{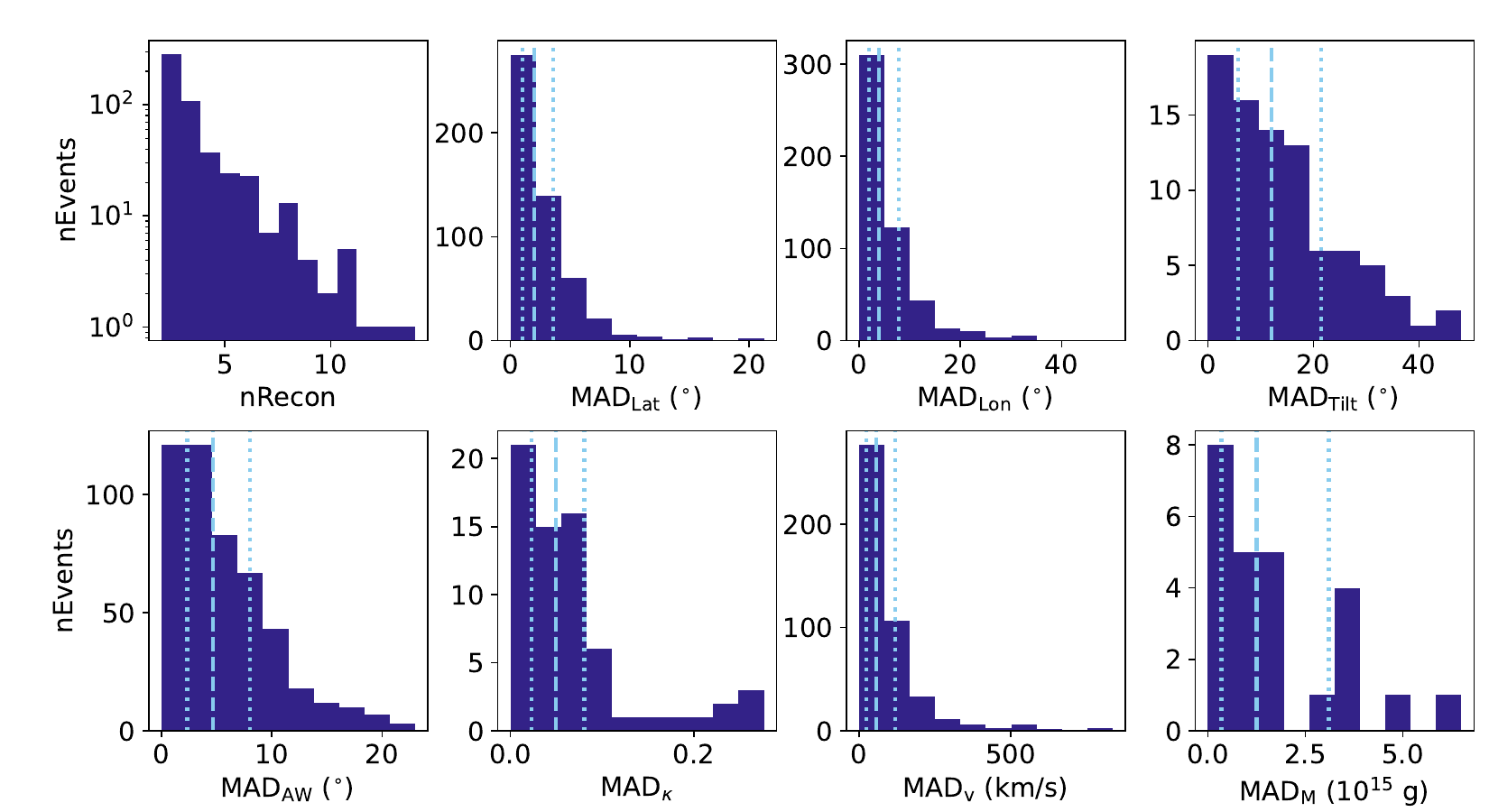}
\caption{Histograms of the MAD for each property within the multi-cat cases. The vertical lines show the median of the MADs (solid line) and the 25th and 75th percentiles of the distributions (dotted lines).}\label{histospread}
\end{figure}

While it is useful to reduce the results of the entire catalog into a few instructive numbers describing the uncertainty, this allows for only a surface understanding and no insights into any trends in uncertainty. To start unpacking these numbers, Figure~\ref{histospread} shows histograms of the MADs for each CME within the multi-cat set. The top left panel shows the number of reconstructions, the same as the top left panel of Figure~\ref{histomeds}, but now we only include the multi-cat cases and show the vertical range on a logarithmic scale. The rest of the panels use a linear y-axis and within them the dashed light blue line represents the median of the distribution and the dotted blue lines the 25th and 75th percentiles.

All of the parameters have histograms peaking toward low MADs. The latitude, longitude, and velocity all have sharp peaks at low values with only a few outliers at slightly higher values. This suggests that for most events the reconstructions are close to the median value and largely consistent with one another, and we can feel relatively confident about our ability to reconstruct these values. However, there are instances where we see fairly large disagreements within a single event. For example, the 2011-02-26T17:08:15 CME appears in both AFFECTScat and AFFECTSgcs but no other catalogs. We use it as an example as the reconstructions were performed by the same AFFECTS team and they associated these reconstructions with the same event. The reconstructed longitudes are 146.2$^{\circ}$ and $-139.1^{\circ}$ for the CAT and GCS versions, respectively, which is a separation of 74.7$^{\circ}$ and a MAD of 37.4$^{\circ}$. This is a back-sided event, which tends to make reconstruction more difficult. STEREO, however, was located at roughly $\pm$90$^{\circ}$ from Earth at this time, which should be favorable for the quality of the reconstruction. We cannot say why these reconstructions differ so much, though we do note that a missing negative sign would bring them to much better agreement. Only 7 events in our sample have a MAD greater than 30$^{\circ}$ in the longitude, suggesting that while these extremely uncertain cases do exist, they are fairly rare.

The AW and tilt distributions are not as sharply peaked, suggesting that these parameters are slightly more difficult to measure precisely than the latitude, longitude, and velocity. We do not see a noticeable separate population of outliers for either property, but this is largely due to there being a finite range of plausible or possible values. In theory, the AW can vary between 0$^{\circ}$ and 180$^{\circ}$ since it is a half-width and the maximum value would correspond to a physically unrealistic complete loop. More often, maximum values tend to be around 90$^{\circ}$, which would correspond to a large CME covering a full hemisphere. Looking at the distributions in Figure~\ref{histomeds}, the LLAMACoRe AW distribution peaks around 30$^{\circ}$ with only a small fraction of cases above 50$^{\circ}$. Only 24 of the events have an AW MAD above 15$^{\circ}$ because the most common CMEs are not big enough to exhibit that large of a spread in the reconstructions. The MADs for the tilt are similarly limited but by the range of possibilities rather than the range of commonly observed values. There is only a 180$^{\circ}$ range in tilts, and it is circular, so the maximum two values can differ by is 90$^{\circ}$. For two reconstructions for a single event, a difference of 90$^{\circ}$ corresponds to a MAD of 45$^{\circ}$, which is the maximum value we observe. This means we have cases, albeit relatively few, where the reconstructed tilts are differing by as much as physically possible. Most cases are better than this but there are times when the reconstructed tilt is completely unreliable.

The distribution of $\kappa$ shows a defined peak toward lower MADs but also a significant secondary population of higher values. We note the small number of events with multiple reconstructions of $\kappa$, the secondary peak of MADs greater than 0.2 corresponds to only 5 events. The histogram in Figure~\ref{histomeds} shows that most CMEs are reconstructed with $\kappa$ between 0.2 and 0.5. For these higher MAD cases, one catalog (always either AFFECTS or KINCAT) uses a $\kappa$ in the range of 0.6-0.9 for the reconstruction. As with most of the other parameters, there is typically a reasonable consensus on the reconstructed $\kappa$, but for a small percentage of events, we find a drastic disagreement.

The bottom right panel of Figure~\ref{histospread} shows the MAD in the mass. Measuring the mass is difficult due to a variety of reasons, including assumptions about where the mass exists relative to the plane of the sky and potential contamination from background objects such as coronal streamers. For a more thorough discussion of these issues we refer the reader to \citeA{Plu19}. We have very few cases with multiple reconstructions of the mass so the distribution is not at all continuous. From the limited information, it appears that the mass reconstructions MADs have similar behavior to the AW with a peak at low values but a relatively broad distribution. This cannot be confirmed, however, without additional data.

\subsection{Trends in the MAD Values}
\begin{figure}[ht!]
\includegraphics[width=\textwidth]{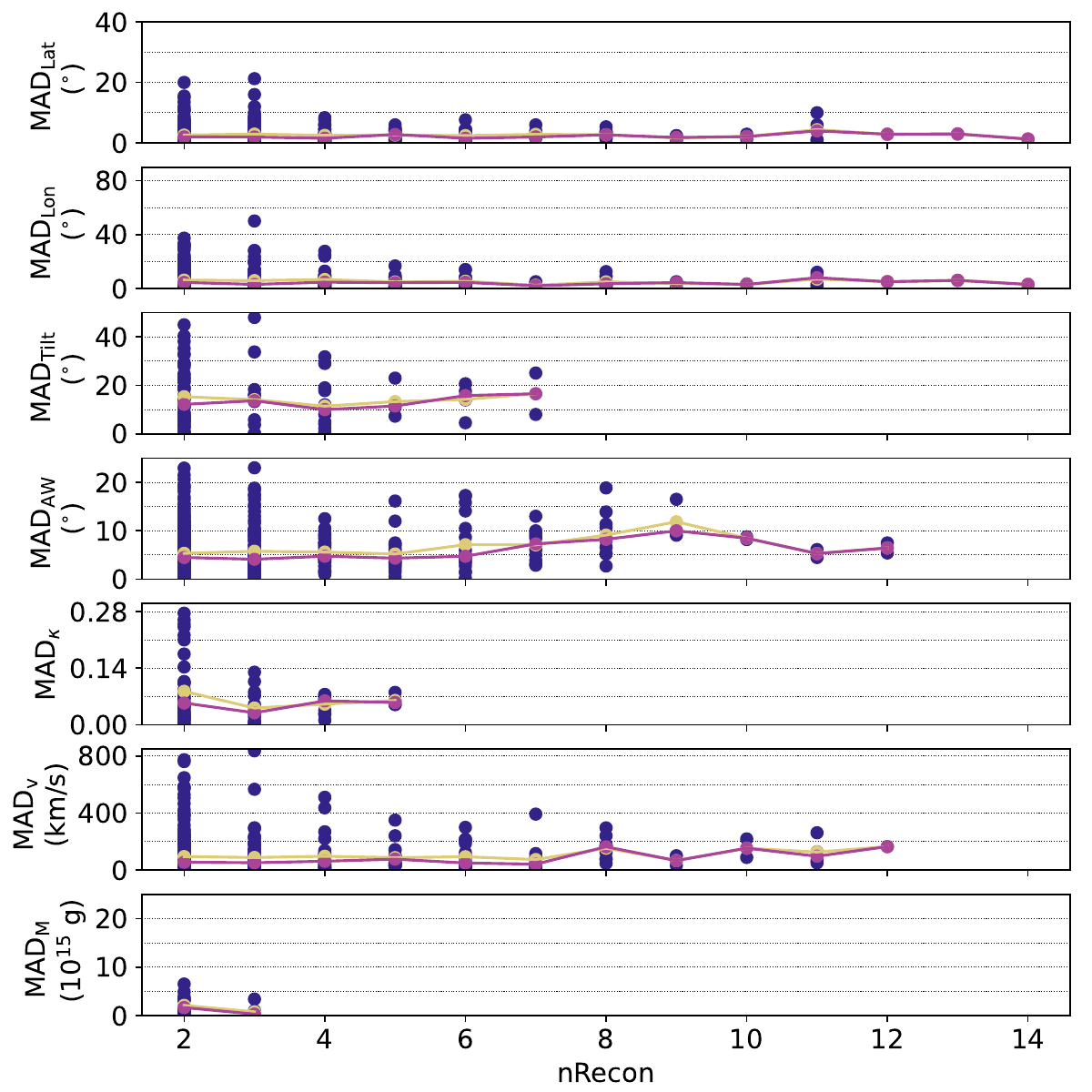}
\caption{MAD of properties for each multi-cat case versus the number of reconstructions. The yellow shows the median and the pink shows the median within each set of events with the same number of reconstructions.}\label{nrecon}
\end{figure}

We look to see if there are any trends in the MAD with the number of reconstructions performed for an event. Figure~\ref{nrecon} includes each CME once as a blue dot and shows the MAD versus the number of reconstructions (nRecon) for that property. We note that there is significant overlap in the dots for the low numbers of nRecon. For each nRecon, we determine the mean and median of the MADs with that number of reconstructions, and show these values in yellow (mean) and pink (median). For the most part, we see little to no dependence of any MAD on the number of reconstructions, the pink and yellow lines are essentially horizontal for most parameters. 

We tend to see larger outliers in the MAD for the low number of reconstructions. The fewer the number of events the more significant the effect a single reconstruction has on the median value and therefore the MAD. Small sample size effects also explain the slight increase in the overall mean/median (yellow/pink lines) toward higher nRecon, particularly for the tilt and the AW. There are only two events that have six reconstructions of the tilt (2008-12-12T08:22:20 and 2010-04-03T11:08:15) and both of them have a fairly wide range in their reconstructed values. We propose that these small sample effects would disappear if we had more events with a large number of independent reconstructions, but cannot confirm this without additional measurements. In general, we find no evidence to support any trend in the most probable MAD versus the number of reconstructions, but the possibility of a large MAD does increase when there are only a few reconstructions.

\begin{figure}[ht!]
\includegraphics[width=\textwidth]{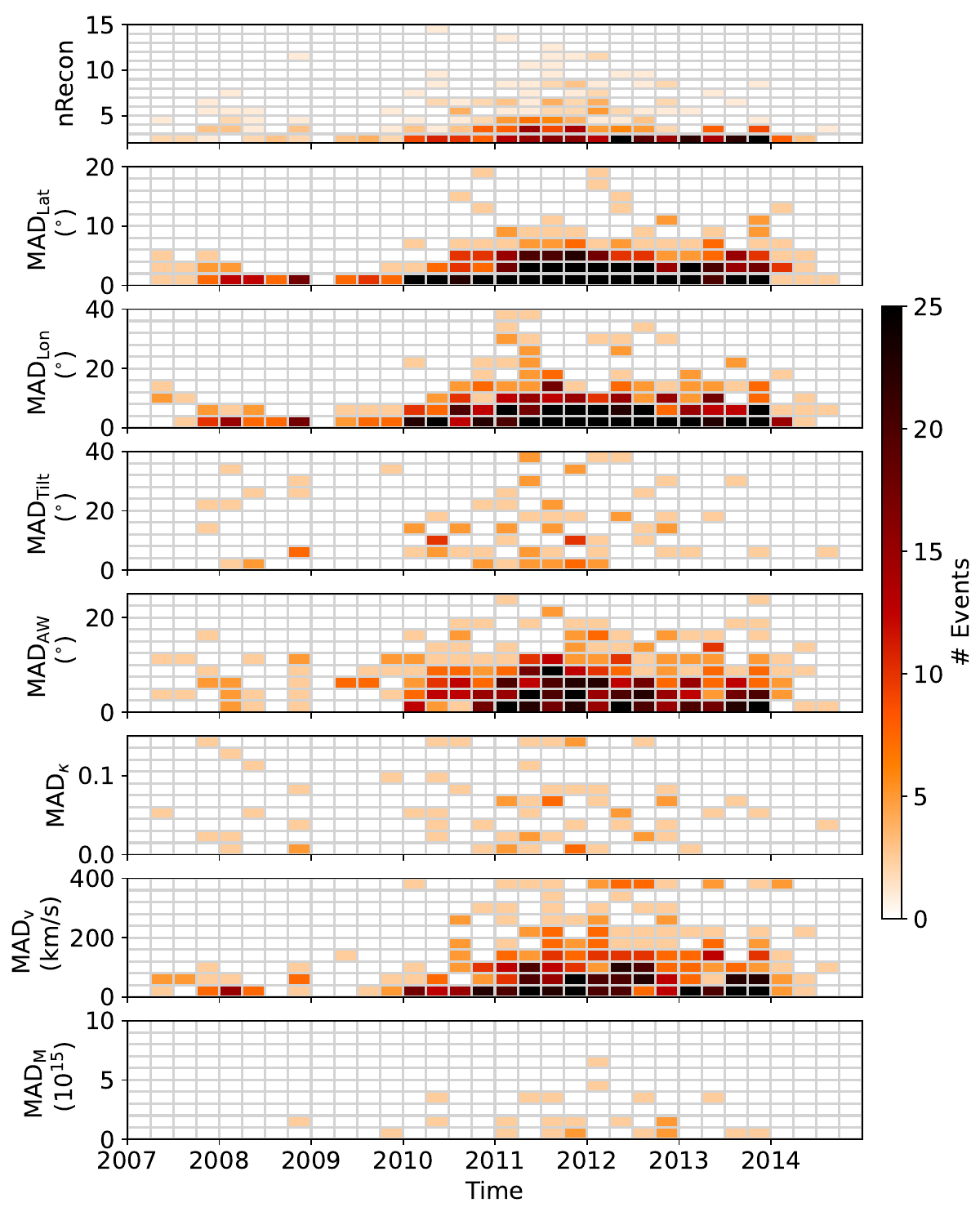}
\caption{Heat map showing the distribution in the MAD of properties as a function of time.}\label{timeline}
\end{figure}

We next investigate how the MAD values change over time, seeking any variations that could be attributed to the solar cycle or satellite separation. Figure~\ref{timeline} shows a heat map of the MAD over time, analogous to Figure~\ref{timelinemedians} but for the uncertainties and with the top panel showing the number of reconstructions per event as opposed to the sunspot number.

We see that the most-populated grid cells stay at about the same MAD values over time. More CMEs occur toward solar maximum, but their most probable uncertainty is the same as the rest of the solar cycle. Toward solar maximum, we do see a larger spread in the range of measured MADs with more cases at high values. This may simply result from the fact that we have a much larger sample of events at this time, or it may be an inherent property of the solar maximum events themselves. At solar maximum, we expect there to be a larger number of very energetic events with larger size and faster speed. If our uncertainties in reconstructions are generally percentage errors, then we would expect larger uncertainties for the more energetic events. Additionally, near maximum, the higher CME occurrence rate may lead to overlapping signatures in the projected white light events. This will make it difficult to accurately separate which feature belongs to the CME of interest versus other events or the background coronal structures.  We cannot distinguish between these effects options with Figure~\ref{timeline} alone, but we can compare the MAD with the median CME properties for each event to gain some insights.

\begin{figure}[ht!]
\includegraphics[width=\textwidth]{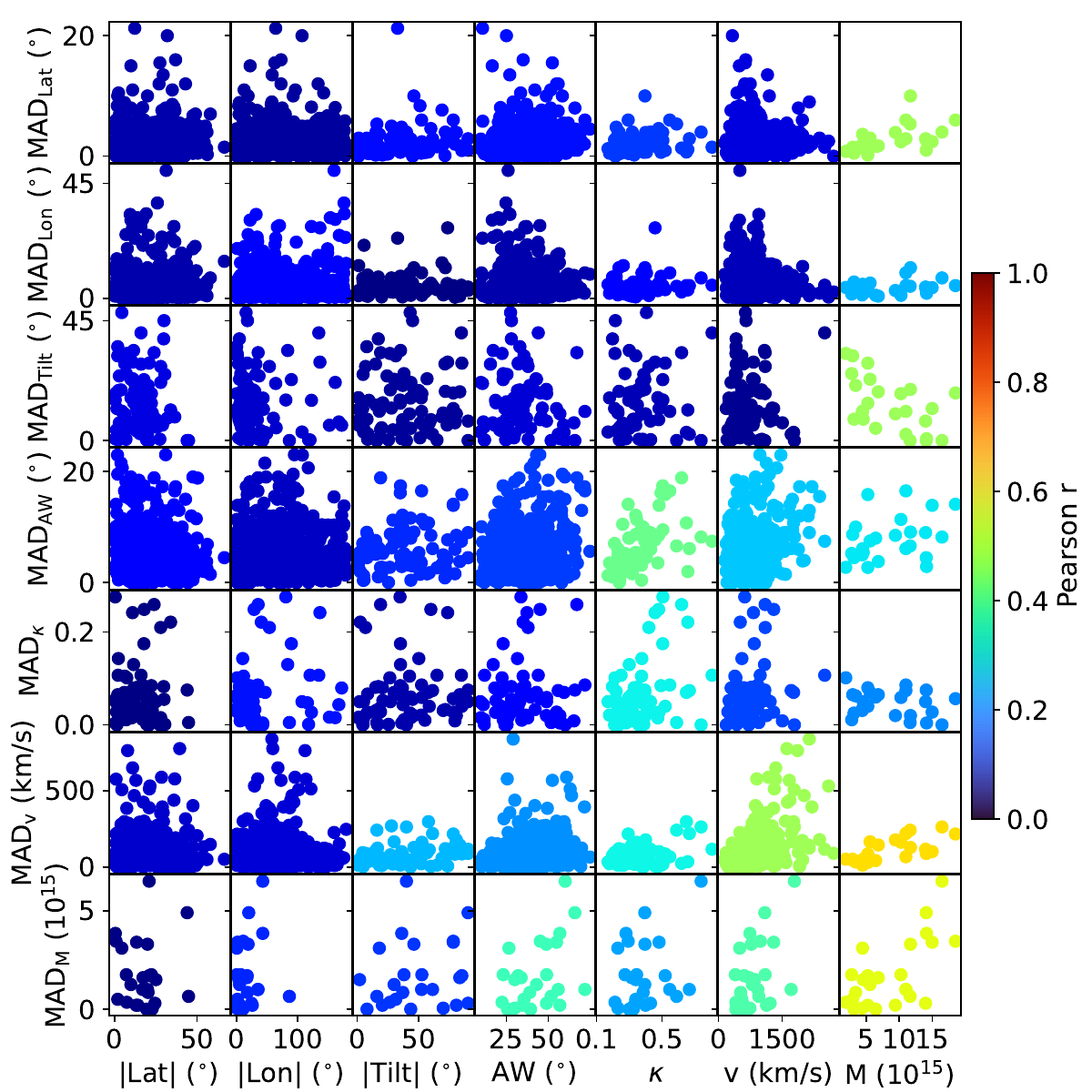}
\caption{Correlations between the MAD of properties versus the median value of properties. Each panel is colored according to the Pearson r-value for that pair of standard deviation and median values.}\label{corrMAD}
\end{figure}

Figure~\ref{corrMAD} shows the scatter of the MADs versus each CME property. Each dot represents a single event and we use the median values for the CME properties. With this analysis, we seek any systematic causes leading to higher uncertainty in the CME reconstructions. Some trends we might expect, such as larger uncertainty in the speed when the speed is higher, but it could also unveil some unexpected patterns. Within Figure~\ref{corrMAD} the individual panels are colored by the correlation coefficient (Pearson r value) corresponding to that MAD and CME parameter.

We compare with the absolute value of the latitude and longitude as we expect there could be a trend with the location relative to the Sun--Earth line but probably no east--west or north--south asymmetry in the uncertainty. Similarly, we might expect variation with high or low inclination, but not whether the tilt is in the clockwise or counterclockwise direction. We have confirmed that such asymmetries with the sign of the position and orientation do not exist, but do not include a figure showing such.  Looking at Figure~\ref{corrMAD}, we actually see no significant correlations with the position and orientation at all. This suggests that, at least while the STEREO probes were separated from Earth, we should be able to reconstruct the position and orientation of CMEs from all source locations with equal uncertainty.

We see some moderate correlations for parameters we might expect to scale with the CME energetics. The MADs for both the velocity and mass increase as the actual velocity and mass increase. There is significant scatter about the general trend, leading only to moderate correlations. We only see a moderate correlation between the AW median and MAD, and interestingly we find a moderate correlation between $\kappa$ and the MAD for the AW. This may hint at a difficulty in matching the wireframe reconstruction shape to the observations when it becomes too fat, but we cannot confirm this at this time.

Figure~\ref{corrMAD} shows an unexpected moderate trend between the CME mass and the MAD for the tilt with higher masses being less uncertain. We only have 25 events in the multi-cat set that have a reconstructed mass so this correlation may just be an artifact of the small sample size. More massive CMEs should have more prominent white light signatures, which should make it easier to reconstruct the CME. This should reduce the uncertainty, but we have no suggestion as to why it would only improve the uncertainty in the tilt but no other parameters. We do note that the MADs for latitude, longitude, $\kappa$, and v tend to be smaller for the subset of the multi-cat set that does have a reconstructed mass than they are for the full set (comparison of the vertical extend of the points within the right-most column panels to any other column). This likely indicates that the cases for which mass reconstructions have been performed tend to be easier to analyze in general, which is why the mass reconstruction could be performed and the rest of the uncertainties are generally low for these events.

Overall, we see very little in terms of correlations between the CME parameters and the uncertainties beyond a slight suggestion of the velocity and mass having an uncertainty proportional to their magnitude. We do not show it but we have also compared the uncertainties to one another and looked for trends there (e.g., tend to have high uncertainty in tilt when we have high uncertainty in longitude) but found nothing noteworthy. We cannot find any systematic variations that could help identify when we expect reconstructions to be more accurate, it seems to be a nonlinear process combining the effects of many different factors.

\section{Discussion}\label{disc}
In combining the existing catalogs into LLAMACoRe we hope to create the most accurate collection of reconstructed CME parameters over the given time span. We certainly feel it is the most thorough collection, given that it includes more events than any of the individual catalogs. We have no way of validating the accuracy of any reconstruction since there is no method to directly measure the ``true'' values of CME properties, everything requires a model making some sort of assumption or approximation. Beyond that, real CMEs are not the rigid, symmetric, idealized structures we use in reconstructions, but rather tenuous structures, so using a single number to describe a property, such as the latitude, may accurately represent the average position but does away with any of the details needed to more-realistically describe the structure. 

In comparing independent reconstructions of the same event we expect slight variations depending on how each particular person fitted a CME. Most events are complicated and it can be difficult to disentangle what portion of the white-light imagery actually corresponds to the CME of interest versus other CMEs or background solar structures such as streamers. If two observers pick slightly different white-light features to fit then we expect there to be a small amount of variation in the reconstructed parameters. If one observer, however, accidentally includes an incorrect white light feature, such as a streamer region, then we expect a much larger variation in the reconstructions. Additionally, if a CME is propagating fast enough, a CME-driven shock may be apparent in the white light images. While reconstructing the shock is also of scientific interest, it is critical to reconstruct it separately from the CME. Distinguishing between these features can be difficult, particularly if one has less experience and is just beginning to perform routine reconstructions.

Visual comparison can show that some reconstructions are clearly better than others, but there are wide ranges of ``good'' parameter space where the reconstruction looks reasonable for the white-light signature and cannot be further constrained through visual comparison alone. Ideally, all reconstructions would fall within this good region and by combining multiple reconstructions we could average out some of the personal preferences within the individual reconstructions and create a new set of best-constrained CME properties. We expect that the vast majority of the reconstructions are reasonable, but that there are probably a few bad cases, either due to mistakes in the reconstruction process or unintentionally introduced human errors. We have no true values to allow for a numerical comparison, nor the time to individually reproduce and compare each of the nearly 3000 reconstructions that are a part of LLAMACoRe. In general, we believe that the individual catalogs are mostly accurate and the derived LLAMACoRe values should be of high quality. We will continually update the online version of LLAMACoRe, updating any reconstructions if errors are identified and adding new cases as they become available.

While we believe the best-constrained properties to be reasonably accurate, we note that the tilt is probably the most suspect. It may be worth examining the individual reconstructions for an event before blindly using our best-constrained value. In the cases where the individual reconstructions are all in general agreement, say within 45$^{\circ}$, then our reported median value does reflect the consensus. However, we know there are cases where the individual reconstructions span the full range of possible tilts. If half the reconstructions suggest high inclination and half suggest low inclination, the best-constrained value will be a mid inclination. Mathematically, this is in the middle of the individual reconstructions, but the spread for this case is so large we really cannot declare anything with any certainty. This is really only an issue for the tilt and not any other reconstructed parameter. The tilt is known to be difficult to reconstruct, but it is a critical parameter for making space weather predictions. Forecasting the southward magnetic field is one of the most important factors in determining the severity of a near-Earth CME impact. In fact, many works have focused on connecting CME axial orientations estimated at the Sun and subsequently at 1~au \cite<e.g.,>{Mar23, Pal22sanchita, Pal17}, finding often striking discrepancies between the two tilts \cite<e.g.,>{Pal18, Xie21, Yur07}. Even more so, it has been proposed that (at least some) CMEs propagate through interplanetary space as significantly kinked structures, and the axis orientation encountered at Earth (or any other point in the heliosphere) only reflects the local conditions of the portion of the cloud that is sampled by a spacecraft \cite{Bot17}. If we do not know the orientation of a CME, how can we know the orientation of its internal magnetic field? We suggest that our inability to accurately reconstruct the orientation of CMEs will be a major limiting factor in space weather predictions in the coming decades and the community should focus on addressing this knowledge gap. 

We have investigated the presence (or lack thereof) of any trends in the magnitude of the uncertainties with respect to the satellite locations. We considered a 2D scatter plot (not shown) with each point colored by the MAD, with the x-axis showing the absolute longitudinal distance between the CME direction and the longitude of STEREO-A, and the y-axis showing the same for STEREO-B. No strong patterns were present. We may see a slight hint of higher uncertainties in the latitude when the CME is moving directly toward one satellite and directly away from the other, but this is likely not statistically significant. Overall, we find that any effects of the relative viewing angles are overwhelmed by the other factors involved in determining the uncertainties.

\section{Conclusion} \label{sec:conc}
We have presented LLAMACoRe, a new meta-catalog that combines existing catalogs into the most extensive collection of 3D CME reconstructions over the prime STEREO era between 2007--2014. LLAMACoRe contains 2954 reconstructions for 1862 different CMEs. Each reconstruction includes, at a minimum, the coronal latitude and longitude of that event. Many reconstructions also include the tilt, angular width, GCS shape parameter (or aspect ratio), velocity, and mass of the CME.

LLAMACoRe contains 511 events with multiple reconstructions, for which we determine the median properties. We use the median properties, in addition to the single-reconstruction events, and analyze the variation in CME properties over time. These results reproduce the known solar cycle variations in CME behavior \cite<e.g.,>{Yas04, Lam19, Gop20}. We see a higher rate of CME occurrence toward solar maximum and a larger number of highly energetic events. We derive new relations between the CME mass, angular width, shape, and velocity, which could be useful for estimating input values for CME forecasting models.

We compare the range in the reconstructions for each event and use their variance to better establish the uncertainty in our reconstruction techniques. We have no measure of the true CME values, so we report the uncertainty as the typical difference we would expect between two independent reconstructions of the same event. We find uncertainties of 4.0$^{\circ}$ in the latitude, 8.0$^{\circ}$ in the longitude, 24.0$^{\circ}$ in the tilt, 9.3$^{\circ}$ in the angular width, 0.1 in the shape parameter $\kappa$, 115~km/s in the velocity, and 2.5$\times$10$^{15}$~g in the mass. Some of these uncertainties can also be expressed as percentages: 27\% for the angular width, 29\% for $\kappa$, 19\% for the velocity, and 38\% for the mass.

We look for any trends in the uncertainty to see if there are instances in which the community is particularly good or bad in performing consistent reconstructions. Unfortunately, we see very little evidence of any useful relations. We do find more outliers with large uncertainty toward solar maximum. This likely results from a combination of more energetic CMEs and the reconstructions being more difficult to perform as the CME occurrence rate increases and cases begin overlapping in the projected fields of view. The most probable uncertainties, however, tend to remain around the same values over the solar cycle. 

We have made our best attempt to incorporate any existing catalogs into LLAMACoRe, but recognize we may have potentially missed some invite any authors to contact us and we will update the online version. We will keep watch over the literature and automatically add any results from new publications with 5 or more reconstructed CMEs. We are happy to include smaller lists if they are sent directly to the LLAMACoRE team, but it is beyond our capabilities to automatically keep track of every publication with a single reconstruction.

We also invite collaboration from the community to help further develop LLAMACoRe as a living repository for any measurements that have been made for observed CMEs. We initially limited the focus to coronal geometric/kinematic reconstructions between 2007--2014 to get the project running, but would like LLAMACoRe to grow in whatever direction the community sees fit. This includes expanding the time range and incorporating additional information such as source regions, associated flare/dimming properties, reconstructions based on heliospheric imaging, and in situ associations/properties.

We propose that the current version of LLAMACoRe has several uses for the community. First, the derived uncertainties should be considered when reporting any CME reconstructions in other works. While our numbers are similar to previously reported uncertainties, they have been derived from the largest number of cases thus far. Second, LLAMACoRe is the most thorough collection of stereoscopic coronal CME reconstructions between 2007--2014. This makes it an excellent starting point for any event case studies, collecting the information from the literature into a single location. This will quickly show whether or not there is consensus on a specific reconstruction. Well-behaved, consensus events would be the most suitable for initially testing new models whereas complicated, highly uncertain events could be more interesting for in-depth observational analyses. When LLAMACoRe grows to include information about source regions and in situ counterparts the amount of information readily at hand will become even more useful.

Another factor is that we have developed a set of over 500 events with relatively well-constrained parameters derived from multiple reconstructions. We propose that using the best-constrained parameters as inputs could be a systematic method for benchmarking interplanetary propagation models, particularly once the CMEs' in situ counterparts have been added to LLAMACoRe. The community has also embraced more machine learning and artificial intelligence approaches over the previous decade \cite<e.g.,>{Cam19}, but appropriately applying these techniques requires larger data sets than often available. The full data set (over 1800 cases), and to a lesser extent the multi-cat set, begin approaching what is reasonable for some advanced techniques. With such a large set there are certainly aspects to explore that the authors have not yet thought of and we invite the community to make use of LLAMACoRe however they see fit.

\section{Open Research}
The present version of LLAMACoRe and future updates to the meta-catalog can be accessed at \href{https://osprei.space/llamacore}{osprei.space/llamacore}. The full data set and the best-constrained properties for each CME can be found in the supplementary information and has been archived at Zenodo at \href{https://doi.org/10.5281/zenodo.10462800}{https://doi.org/10.5281/zenodo.10462800}.
Smoothed sunspot data were taken from the World Data Center SILSO, Royal Observatory of Belgium, Brussels, available at \href{https://www.sidc.be/SILSO/datafiles}{sidc.be/SILSO/datafiles}.

%AGU requires an Availability Statement for the underlying data needed to understand, evaluate, and build upon the reported research at the time of peer review and publication.

%Authors should include an Availability Statement for the software that has a significant impact on the research. Details and templates are in the Availability Statement section of the Data and Software for Authors Guidance: \url{https://www.agu.org/Publish-with-AGU/Publish/Author-Resources/Data-and-Software-for-Authors#availability}

%It is important to cite individual datasets in this section and, and they must be included in your bibliography. Please use the type field in your bibtex file to specify the type of data cited. Some options include Dataset, Software, Collection, ComputationalNotebook. Ex: 

%@misc{https://doi.org/10.7283/633e-1497,
%  doi = {10.7283/633E-1497},
%  url = {https://www.unavco.org/data/doi/10.7283/633E-1497},
%  author = {de Zeeuw-van Dalfsen, Elske and Sleeman, Reinoud},
%  title = {KNMI Dutch Antilles GPS Network - SAB1-St_Johns_Saba_NA P.S.},
%  publisher = {UNAVCO, Inc.},
%  year = {2019},
%  type = {dataset}
%}

%%%%%%%%%%%%%%%%%%%%%%%%%%%%%%%%%%%%%%%%%%%%%%

\acknowledgments
First and foremost we acknowledge the authors and teams responsible for compiling the individual catalogs that comprise LLAMACoRe. None of this work would have been possible without these existing sources. The AFFECTS STEREO/SECCHI/COR2 CME catalog is generated and maintained at the Institute for Astrophysics of the University of Goettingen, supported by the German Space Agency DLR and the European Union in collaboration with the U.S.\ Naval Research Laboratory, Washington. We acknowledge the Community Coordinated Modeling Center (CCMC) at Goddard Space Flight Center for the use of the DONKI catalog. The HELCATS KINCAT catalog is generated and maintained at the Institute for Astrophysics of the University of Goettingen, supported by the German Space Agency DLR and the European Union. The remaining catalogs were pulled from the online published papers and we acknowledge the work of those authors and the journals that host the content.\\
C.~Kay is supported by the National Aeronautics and Space Administration under Grant 80NSSC19K0909 issued through the Heliophysics Early Career Investigators Program. E.~Palmerio acknowledges support from NASA's Heliophysics Guest Investigators-Open program (grant no.\ 80NSSC23K0447).

%% ------------------------------------------------------------------------ %%
%% References and Citations

%%%%%%%%%%%%%%%%%%%%%%%%%%%%%%%%%%%%%%%%%%%%%%%
%
% \bibliography{<name of your .bib file>} don't specify the file extension
%
% don't specify bibliographystyle

% In the References section, cite the data/software described in the Availability Statement (this includes primary and processed data used for your research). For details on data/software citation as well as examples, see the Data & Software Citation section of the Data & Software for Authors guidance
% https://www.agu.org/Publish-with-AGU/Publish/Author-Resources/Data-and-Software-for-Authors#citation

%%%%%%%%%%%%%%%%%%%%%%%%%%%%%%%%%%%%%%%%%%%%%%%

%\bibliography{master}

%Reference citation instructions and examples:
%
% Please use ONLY \cite and \citeA for reference citations.
% \cite for parenthetical references
% ...as shown in recent studies (Simpson et al., 2019)
% \citeA for in-text citations
% ...Simpson et al. (2019) have shown...
%
%
%...as shown by \citeA{jskilby}.
%...as shown by \citeA{lewin76}, \citeA{carson86}, \citeA{bartoldy02}, and \citeA{rinaldi03}.
%...has been shown \cite{jskilbye}.
%...has been shown \cite{lewin76,carson86,bartoldy02,rinaldi03}.
%... \cite <i.e.>[]{lewin76,carson86,bartoldy02,rinaldi03}.
%...has been shown by \cite <e.g.,>[and others]{lewin76}.
%
% apacite uses < > for prenotes and [ ] for postnotes
% DO NOT use other cite commands (e.g., \citet, \citep, \citeyear, \citealp, etc.).
% \nocite is okay to use to add references from your Supporting Information
%

\end{document}